\documentclass[11pt,a4paper]{article}
\pdfoutput=1 

\usepackage{jheppub}

\usepackage[utf8]{inputenc}
\usepackage[T1]{fontenc}
\usepackage{lmodern}
\usepackage{fontawesome}
\usepackage{hyperref}
\usepackage{tikz}
\usepackage{subcaption}
\usepackage{mathtools}
\usepackage{xcolor}
\usepackage{amsmath,amssymb}
\usepackage{booktabs}
\usepackage{bbding}
\usepackage{graphicx} 
\usepackage{tikz-feynman}
\usepackage{bm}
\usepackage{multirow}
\usepackage{placeins}
\usepackage{float}

\newcommand{\OO}{\ensuremath{\mathcal{O}}}
\newcommand{\Op}[1]{\OO_{\sss #1}}
\newcommand{\sss}{\scriptscriptstyle}
\newcommand{\pdp}{\ensuremath{\varphi^\dagger\varphi}}
\newcommand{\ccc}[3]{c_{#2}^{#1 (#3)}}

\def\lra#1{\overset{\text{\scriptsize$\leftrightarrow$}}{#1}}

\usepackage{graphicx}
\usepackage{bm}
\usepackage{hyperref}
\hypersetup{
  colorlinks=true,
  linkcolor=blue,
  citecolor=blue,
  urlcolor=blue
}

\title{
  Exploring the SMEFT landscape: Bayesian Model Selection for indirect discovery
}

\author{Luca Mantani}

\affiliation{Instituto de Física Corpuscular (IFIC), Universidad de Valencia--CSIC, E-46980 Valencia, Spain}

\abstract{
We develop a framework for indirect discovery in the Standard Model Effective Field 
Theory (SMEFT) based on Bayesian model selection over operator subsets. We argue 
that SMEFT should be understood as a structured space of competing hypotheses rather 
than a single high-dimensional model, with each operator subset corresponding to a 
physically distinct low-energy realisation of new dynamics. Bayesian inference 
is applied at the level of model space itself, assigning posterior probabilities to 
operator subsets and marginal inclusion probabilities to individual operators. A genetic algorithm 
efficiently navigates the high-dimensional discrete model space, concentrating 
evaluations in the high-posterior region, while the Bayesian Information Criterion 
provides a tractable approximation to the Bayesian evidence. We apply this framework to a dataset comprising electroweak precision observables 
from LEP and Higgs, top-quark, and diboson measurements from LHC Run~2, at both 
linear and quadratic order in the Wilson coefficients, with one-loop renormalisation 
group evolution systematically included. The analysis finds no statistically 
significant evidence for any departure from the SM, and demonstrates that Bayesian Model Average 
posteriors on Wilson coefficients carry substantially improved characterisation 
potential compared to traditional global fits. The operator correlation matrix 
encodes the relational structure of the model posterior, identifying operator pairs 
that co-appear in high-posterior models and flat directions where additional 
measurements would be most valuable. The sensitivity of all results to the choice 
of matching scale $\mu_0$ is assessed, and its promotion to a continuous parameter 
of inference is identified as a natural extension of the framework.
}

\begin{document}

\maketitle

\section{Introduction}

The Standard Model (SM) of particle physics stands as one of the most precisely tested 
theories in science, with its predictions confirmed across an extraordinary range of energy 
scales. Its most recent triumph, the discovery of the Higgs boson at the Large Hadron 
Collider (LHC) in 2012~\cite{ATLAS:2012yve, CMS:2012qbp}, completed the particle content of the theory and validated the 
mechanism of electroweak symmetry breaking. Yet, despite this remarkable success, the SM 
is widely understood to be incomplete. A number of observational facts, the existence 
of dark matter~\cite{Cirelli:2024ssz}, the baryon asymmetry of the universe~\cite{Bodeker:2020ghk}, and the non-zero masses of neutrinos~\cite{King:2003jb}, find no explanation within its framework. 
Equally compelling are purely theoretical 
shortcomings, such as the hierarchy problem and the strong CP problem~\cite{Craig:2022eqo}, which suggest that 
the SM must be regarded as an effective description valid up to some finite energy scale, 
beyond which a more fundamental theory takes over.

The LHC was conceived in part to address this situation directly, with high expectations 
that new degrees of freedom at the TeV scale would emerge and shed light on both the 
observational and theoretical tensions. While this direct discovery program remains ongoing 
and the possibility of new particle discoveries is not excluded, the absence of clear 
signals of physics beyond the SM in Run~1 and Run~2 data has progressively shifted the 
landscape. It is increasingly plausible that new physics (NP) resides at energy scales beyond 
the direct reach of present colliders, in which case the experimental strategy must evolve. 
The focus shifts from direct production of new states to indirect searches, seeking subtle 
but systematic deviations from SM predictions in precision observables. The forthcoming 
High-Luminosity LHC (HL-LHC) programme~\cite{Cepeda:2019klc, Azzi:2019yne}, with its substantial increase in integrated 
luminosity and corresponding reduction in statistical uncertainties, represents a critical 
opportunity to advance this indirect search programme.

In this context, the Standard Model Effective Field Theory (SMEFT)~\cite{Isidori:2023pyp} has emerged in 
recent years as the framework of choice for indirect searches of NP. The SMEFT 
extends the SM under the assumption of heavy new dynamics and a linear realisation of 
electroweak symmetry breaking, providing a systematic and model-independent 
parametrisation of possible deviations from SM predictions. New interactions are 
suppressed by powers of the NP scale $\Lambda$, enabling a natural power 
counting that organises corrections by their canonical mass dimension. In practice, most 
analyses retain operators up to dimension six~\cite{Grzadkowski:2010es}, which represent the leading corrections 
under the assumption of lepton number conservation. 
Dimension-eight operators~\cite{Murphy:2020rsh} have also been considered in the 
literature~\cite{Boughezal:2022nof, Dawson:2021xei, Corbett:2023qtg}, though the natural expectation is that dimension-six effects dominate 
and dimension-eight contributions become relevant only once the former are sufficiently 
well constrained. This hierarchy provides a practical and well-motivated truncation of 
the effective expansion.

Mapping out the SMEFT parameter space poses a daunting challenge. The operator 
content of the theory is vast: depending on the underlying flavour assumptions, the number of 
independent dimension-six operators runs into the hundreds, each associated with a 
Wilson coefficient that must be constrained by data. Meeting this challenge requires 
the simultaneous analysis of a wide array of measurements, drawing on observables from 
qualitatively different processes and spanning a broad range of energy scales. The 
dominant paradigm for tackling this problem has been the global SMEFT 
fit~\cite{de_Blas_2020,Brivio_2022,Ellis_2021,Cirigliano:2016nyn,Cirigliano:2023nol,
Garosi:2023yxg,Bissmann:2019gfc,Bissmann:2020mfi,Allwicher:2023shc,Bruggisser:2021duo,
Bruggisser:2022rhb,Bellafronte:2023amz,Grunwald:2023nli,Hiller:2024vtr,Hiller:2025hpf,terHoeve:2025omu, Mantani:2025bqu,terHoeve:2025gey,Celada:2024mcf,Maura:2025rcv,
deBlas:2025xhe, CMS:2025ugn, ATLAS:2022xyx, Armadillo:2026mvp, Smolkovic:2026cba}, in which a large number of operators are constrained simultaneously 
against a comprehensive dataset. Such analyses are invaluable for establishing general 
bounds on NP, but they come with an inherent limitation from a discovery 
perspective: when a genuine signal occupies only a narrow region of operator space, its 
statistical footprint risks being washed out by marginalisation over the many directions 
in parameter space to which the data are largely insensitive.

In Ref.~\cite{Hirsch:2025qya}, we proposed a data-driven approach based on Bayesian 
model selection that goes beyond traditional global fits, demonstrating how targeted 
operator selection can improve both the discovery and characterisation potential of 
SMEFT analyses and offering a concrete picture of how an indirect discovery of NP could unfold in practice. A related methodology employing the Akaike Information Criterion to reduce the dimensionality of SMEFT fits has been proposed in Ref.~\cite{Cirigliano:2023nol}, though restricted to toggling small groups of operators rather than navigating the full combinatorial space.

In the present work, we formalise and extend this 
framework, advocating for a new paradigm in EFT analyses centred on the systematic 
characterisation of NP signals as they arise from UV completions of the SM. 
The paper is organised as follows. In Sec.~\ref{sec:epistemology} we discuss the epistemological 
foundations of the approach. Sec.~\ref{sec:stat} presents the statistical framework underpinning 
the analysis. The methodology for the present analysis is described in Sec.~\ref{sec:methodology}, and results are presented in 
Sec.~\ref{sec:eft_results}. Finally, we summarise our conclusions and outlook in Sec.~\ref{sec:conclusions}.

\section{The nature of indirect discovery in the SMEFT}
\label{sec:epistemology}

The goal of any search for physics beyond the SM is not only to constrain the parameters of a given theoretical extension, 
but also to determine whether the data provide evidence for such an extension in the first place.

Parameter estimation, the dominant mode of traditional SMEFT analyses, assumes that a model 
has already been selected and asks how its free parameters are constrained by 
data. This is a well-posed and valuable task, but it presupposes an answer to a 
prior question: is the chosen model the right one? In the absence of a clear 
ultraviolet signal pointing to a specific operator structure, this question must 
itself be treated as an object of inference. Discovery is fundamentally a question about models rather than parameters, 
and its natural language is model comparison: whether the data support the 
introduction of additional structure, and if so, which operators are 
implicated.

\subsection{SMEFT as a structured hypothesis space}
\label{subsec:smeft-hyp}

The SMEFT is conventionally presented as a single 
model characterised by a large number of Wilson coefficients. We argue that this framing
does not reflect the actual epistemic situation of a bottom-up EFT analysis.

In a genuinely agnostic, bottom-up approach, no assumption is made about the 
nature of the ultraviolet completion. The ultimate objective is to establish 
which operators are present at all. Different UV completions generate different 
operator subsets, often with specific patterns of correlations dictated by the 
symmetries and field content of the UV theory. Each distinct operator combination 
therefore corresponds to a physically distinct low-energy hypothesis about the nature of NP, 
and in this view SMEFT should be understood as a space of such hypotheses.

This perspective is most naturally articulated at the matching scale 
$\mu_0 \sim M_{\mathrm{NP}}$, where the heavy degrees of freedom are 
integrated out and the presence or absence of each operator directly 
reflects the dynamics of the underlying UV completion.
Below $\mu_0$, renormalisation group 
evolution (RGE) induces operator mixing~\cite{Jenkins:2013wua,Jenkins:2013zja,Alonso:2013hga,Born:2026xkr}, so that coefficients initialised in a sparse 
subset at the matching scale will generate contributions to other operators at 
lower energies. The hypothesis space is therefore most cleanly defined at high 
scales, with the low-energy phenomenology being a derived picture obtained by 
evolving the high-scale operator content down to the experimentally accessible 
regime.

The structure of this hypothesis space is further enriched by the hierarchical 
nature of operator generation. At tree level, a given UV completion generates a 
specific and often limited set of operators~\cite{deBlas:2017xtg}, while additional operators are 
induced at higher loop orders with coefficients suppressed by the corresponding 
loop factors~\cite{Guedes:2023azv, Guedes:2024vuf}. However, this does not rigidly determine the order in which 
operators can be discovered: whether a tree-level or loop-induced operator leaves a 
statistically significant imprint first depends on the interplay between its 
magnitude and the sensitivity of the available data. The effective 
hierarchy in the hypothesis space is therefore a dynamical feature, shaped 
jointly by the structure of the UV completion and by the experimental precision 
frontier.

A conceptually important caveat concerns the basis dependence of operator 
subsets. The decomposition of a given physical effect into a sum of EFT operators 
is not unique: different choices of operator basis, related by field redefinitions, 
can redistribute the same physical content across different operator combinations~\cite{Criado:2018sdb}. 
As a consequence, a hypothesis that corresponds to a single active operator in 
one basis may require a specific combination of several operators in another. 
The implications of this for model selection are discussed in the following section.

\subsection{From parameter estimation to model comparison}

The traditional paradigm for SMEFT analyses is the global fit, in which a large 
number of operators are activated simultaneously and constrained against a 
comprehensive dataset. This approach is well suited to its intended purpose: 
establishing conservative, model-independent bounds on Wilson coefficients under 
the most general SMEFT hypothesis. Whether any particular deformation of the SM 
is actually favoured by the data is a distinct question, and one that requires 
a different set of tools.

The limitations of this approach become apparent in a discovery context.
When many operators are active simultaneously, the posterior over any individual 
coefficient reflects not only the data's sensitivity to that operator, but also 
marginalisation over all directions in parameter space to which the data are 
largely insensitive. In a high-dimensional fit, this 
marginalisation generically broadens coefficient posteriors and dilutes genuine 
signals: a true NP contribution aligned with a narrow direction in operator space 
can be absorbed or obscured by the collective flexibility of the many weakly 
constrained directions surrounding it~\cite{Hirsch:2025qya}. The result is that a global fit may return 
posteriors broadly consistent with zero even when the data contain a statistically 
significant deviation.

The scientific question of interest from a discovery perspective is therefore not 
whether a particular Wilson coefficient is consistent with zero within a global 
fit, but whether its inclusion is warranted by the data at all. Answering this 
requires a framework in which model complexity is explicitly penalised, and in 
which a simpler effective description is preferred unless a more complex one 
delivers a commensurate improvement in fit quality. This is the logic of Bayesian 
model comparison: an operator is favoured because a model containing that operator describes the data 
meaningfully better than one without it, once the cost of the additional degree of 
freedom is accounted for.

A conceptually interesting subtlety concerns the basis dependence of operator 
counting. The decomposition of a given physical effect into EFT operators is not 
unique: different bases, related by field redefinitions, redistribute the same 
physical content across different operator combinations. A hypothesis corresponding 
to a single active operator in one basis can appear in another as a combination 
of operators with fixed coefficient ratios, a one-parameter subspace of a 
nominally multi-operator model. Since complexity penalties count free parameters, 
the penalty is effectively basis-invariant for physically equivalent hypotheses.
What is genuinely basis-dependent is the discretisation of the hypothesis space, 
which hypotheses are even expressible as single on/off switches, 
a feature inherent to any fixed-basis SMEFT analysis. Throughout this work we adopt the Warsaw 
basis~\cite{Grzadkowski:2010es}, the standard choice in the field, and regard a 
systematic understanding of basis dependence in model selection as an important 
direction for future work.

\subsection{The probabilistic language of discovery}

The evidential structure of indirect discovery differs fundamentally from 
that of direct searches, where evidence typically concentrates in a single 
observable and the question of discovery reduces to assessing its local 
statistical significance. Indirect discovery through model selection is 
intrinsically diffuse and cumulative: evidence accumulates gradually across 
datasets and energy scales, distributed over many observables simultaneously, 
and a single measurement is unlikely to be decisive on its own. The Bayesian framework 
captures this naturally, as new data are incorporated, the posterior over 
model space is updated continuously, with genuinely active operators 
accumulating support across observables and degeneracies between competing 
combinations progressively lifted.

This cumulative view has concrete implications for how results should be 
reported and interpreted at each stage of the experimental programme. 
A snapshot of the model posterior at a given moment in time 
represents the current state of evidence, not a final verdict. 
The appropriate scientific output is therefore not a single claim 
about the presence or absence of NP, but a trajectory: 
the evolution of operator inclusion probabilities and the gradual concentration 
of posterior weight as data accumulate.

Correspondingly, the value of a new dataset lies not only 
in its potential to produce a statistically significant excess, but in its 
capacity to resolve existing degeneracies and sharpen inclusion probabilities 
for operators of physical interest. This provides a natural and quantitative 
language for assessing the discovery potential of future experimental facilities~\cite{deBlas:2025gyz}, one 
sensitive to the full structure of the hypothesis space rather than to the 
detectability of any single benchmark signal.

\section{Statistical framework}
\label{sec:stat}

Having established the epistemological foundations of our approach, we now 
develop the formal statistical framework that underlies the analysis. The central 
thesis of the preceding section, that SMEFT should be understood as a structured 
space of competing hypotheses rather than a single high-dimensional model, calls 
for a statistical language capable of assigning probabilities to model structures 
themselves, not merely to parameter values within a fixed model. Bayesian inference 
provides precisely this language\footnote{For a review of Bayesian inference in the context of 
cosmology and particle physics, see Ref.~\cite{Trotta:2008qt}; 
for its foundations as the logic of scientific reasoning, 
see Ref.~\cite{Jaynes:2003jaq}.}, and in what follows we develop its application to 
the SMEFT context in detail, from parameter-level posteriors through to model 
comparison, model averaging, and the practical approximations that make these 
computations tractable at scale.

\subsection{Bayesian inference over the SMEFT model space}
\label{sec:models}

In the SMEFT, the effective Lagrangian truncated at dimension six reads
\begin{equation}
\mathcal{L}_{\mathrm{SMEFT}}
=
\mathcal{L}_{\mathrm{SM}}
+
\sum_{i=1}^{d} \frac{c_i}{\Lambda^2}\, \mathcal{O}_i ,
\end{equation}
where $\mathcal{O}_i$ are independent gauge-invariant dimension-six operators
and $c_i$ their Wilson coefficients, defined within a fixed operator basis of
dimension $d$ determined by the symmetry and flavour assumptions adopted. 
Different UV completions generate different subsets of
operators, each corresponding to a physically distinct low-energy hypothesis
about the nature of NP. As argued in Sec.~\ref{sec:epistemology}, SMEFT is
therefore best understood not as a single model with many parameters, but as
a structured space of competing effective descriptions. This perspective
naturally calls for a statistical language capable of assigning probabilities
to model structures themselves, not merely to parameter values within a
fixed model.

To make this concrete, we encode each hypothesis by a binary vector
\begin{equation}
\vec{b} = (b_1, \ldots, b_d), \qquad b_i \in \{0,1\},
\end{equation}
where $b_i = 1$ indicates that operator $\mathcal{O}_i$ is active and
$b_i = 0$ that it is absent. The full model space is then
$\mathcal{M} = \{0,1\}^d$, with cardinality $2^d$. For a given $\vec{b}$,
the continuous parameters of the model are the active Wilson coefficients
$\vec{c}_{\vec{b}} = \{ c_i \mid b_i = 1 \}$, with all remaining
coefficients set to zero. The Bayesian evidence for model $\vec{b}$ is
obtained by marginalising over these parameters,
\begin{equation}
p(D \mid \vec{b})
=
\int d\vec{c}_{\vec{b}}\;
p(D \mid \vec{c}_{\vec{b}},\, \vec{b})\,
p(\vec{c}_{\vec{b}} \mid \vec{b}),
\end{equation}
where $p(D \mid \vec{c}_{\vec{b}}, \vec{b})$ is the likelihood and
$p(\vec{c}_{\vec{b}} \mid \vec{b})$ the prior over Wilson coefficients.
Traditional global SMEFT fits correspond to conditioning on a single,
high-dimensional choice of $\vec{b}$; the framework developed here
treats $\vec{b}$ itself as an object of inference, and the evidence
$p(D \mid \vec{b})$ as the central quantity through which competing
hypotheses are compared.

Bayesian inference extends naturally from continuous parameters to discrete
hypotheses by treating the model index itself as a random variable. Applying
Bayes' theorem at the level of models yields the posterior over operator
subsets,
\begin{equation}
p(\vec{b} \mid D)
=
\frac{p(D \mid \vec{b})\, p(\vec{b})}
     {\sum_{\vec{b}'} p(D \mid \vec{b}')\, p(\vec{b}')},
\end{equation}
where $p(\vec{b})$ is the prior probability assigned to model $\vec{b}$.
The prior $p(\vec{b})$ encodes any a priori preference over operator subsets 
before the data are examined. In this work we adopt a uniform prior over all 
$2^d$ models, which does not prejudge which operators are a priori more 
plausible and leaves the task of penalising model complexity entirely to the 
likelihood through marginalisation.

The posterior $p(\vec{b} \mid D)$ is the fundamental object of this
framework. Through the marginalisation over Wilson coefficients encoded in
the evidence $p(D \mid \vec{b})$, models whose additional degrees of freedom
are not required by the data are automatically suppressed, a quantitative
realisation of Occam's razor. A direct consequence of model-level inference is the possibility to define marginal probabilities for individual operators,
\begin{equation}
p(b_i = 1 \mid D)
=
\sum_{\vec{b} : b_i = 1} p(\vec{b} \mid D),
\end{equation}
which measures the degree to which the data support the presence of
$\mathcal{O}_i$ after marginalising over all other operator combinations.
Such a notion of operator relevance is not accessible in traditional global
fits, where the operator content is fixed by assumption.

The posterior $p(\vec{b} \mid D)$ addresses the question of model selection:
which operator subsets are statistically favoured by the data. A separate
but related goal is parameter estimation, which is addressed at the level
of the coefficient posteriors $p(\vec{c}_{\vec{b}} \mid D, \vec{b})$,
characterising the values of the Wilson coefficients within each supported
model.
When several models receive comparable posterior support, conditioning
inference on a single best model risks underestimating uncertainties and
over-interpreting fluctuations in both respects. Bayesian model averaging (BMA) resolves this
by weighting predictions by their model posterior.
For any quantity of interest $\Delta$, such as a prediction or a Wilson coefficient, the model-averaged posterior is defined as
\begin{equation}
p(\Delta \mid D)
=
\sum_{\vec{b}} p(\Delta \mid D, \vec{b})\, p(\vec{b} \mid D).
\end{equation}
This expression makes explicit that uncertainty in the underlying effective description is propagated into the final inference.
Only in the limit where a single model dominates the posterior does Bayesian model averaging reduce to inference within that model.

In the SMEFT context, this construction admits a particularly transparent interpretation.
For a Wilson coefficient $c_i$, the unconditional posterior $p(c_i \mid D)$ obtained from model averaging is a mixture distribution.
Models in which the corresponding operator is absent enforce $c_i = 0$, while models in which it is present contribute a continuous posterior.
As a result, the unconditional distribution generically takes the form of a point mass at the origin supplemented by a continuous component,
\begin{equation}
p(c_i \mid D)
=
\bigl(1 - p(b_i = 1 \mid D)\bigr)\,\delta(c_i)
+
p(b_i = 1 \mid D)\, p(c_i \mid D, b_i = 1),
\end{equation}
where $p(c_i \mid D, b_i = 1)$ denotes the coefficient BMA posterior conditional on the operator being present.

While this unconditional posterior is formally correct and essential for fully model-averaged predictions, it can be less transparent for phenomenological interpretation.
In particular, the presence of a point mass at $c_i = 0$ complicates the construction and interpretation of conventional intervals or “bounds” on Wilson coefficients.
For this reason, it is often more informative to report separately the operator inclusion probability $p(b_i = 1 \mid D)$ and the conditional posterior $p(c_i \mid D, b_i = 1)$.
This two-layer characterisation cleanly separates the question of operator existence from that of coefficient magnitude and will play a central role in the presentation of our results.

\subsection{Model selection as a discovery strategy}
\label{sec:discovery}

The framework developed above can be deployed directly as a discovery
tool by posing a more targeted question than the general model comparison
of Sect.~\ref{sec:models}: not which operator subsets are favoured by the
data, but whether the data provide evidence for any departure from the SM
at all. This is a distinct inferential setup, in which the SM is elevated
to the status of a dedicated null hypothesis rather than treated as one
model among $2^d$ on equal footing. It is naturally formalised by
assigning prior probabilities to the SM and to the composite
hypothesis that some SMEFT deformation is realised,
\begin{equation}
p(\mathcal{H}_{\mathrm{SM}}) + p(\mathcal{H}_{\mathrm{NP}}) = 1,
\end{equation}
where $\mathcal{H}_{\mathrm{NP}}$ is itself a mixture over all non-trivial
operator subsets,
\begin{equation}
p(D \mid \mathcal{H}_{\mathrm{NP}})
=
\sum_{\vec{b} \neq \vec{0}} p(D \mid \vec{b})\, p(\vec{b} \mid \mathcal{H}_{\mathrm{NP}}).
\end{equation}
The corresponding Bayes factor,
\begin{equation}
B_{10} = \frac{p(D \mid \mathcal{H}_{\mathrm{NP}})}{p(D \mid \mathcal{H}_{\mathrm{SM}})},
\end{equation}
provides a prior-independent summary of the relative evidence for NP 
over the SM, with the marginalisation over the full SMEFT hypothesis space 
already incorporated by construction. The posterior probability of the SM 
given the data then follows from Bayes' theorem,
\begin{equation}
\label{eq:sm_posterior}
p(\mathcal{H}_{\mathrm{SM}} \mid D)
=
\frac{p(D \mid \mathcal{H}_{\mathrm{SM}})\, p(\mathcal{H}_{\mathrm{SM}})}
{p(D \mid \mathcal{H}_{\mathrm{SM}})\, p(\mathcal{H}_{\mathrm{SM}}) 
+ p(D \mid \mathcal{H}_{\mathrm{NP}})\, p(\mathcal{H}_{\mathrm{NP}})}
=
\frac{1}{1 + B_{10}\, \dfrac{p(\mathcal{H}_{\mathrm{NP}})}{p(\mathcal{H}_{\mathrm{SM}})}},
\end{equation}
which reduces to $(1 + B_{10})^{-1}$ upon adopting equal priors 
$p(\mathcal{H}_{\mathrm{SM}}) = p(\mathcal{H}_{\mathrm{NP}}) = 1/2$.
This is the natural quantity to report as a discovery statistic: a direct 
probability statement about the SM in light of the data, with the prior 
dependence fully transparent. The two setups are complementary: the uniform 
prior over all $2^d$ models is the natural choice for exploring the full 
structure of the model posterior, while the SM vs.\ NP comparison compresses 
this into a single targeted discovery statistic.

To calibrate the strength of evidence, we adopt the Jeffreys 
scale~\cite{Jeffreys:1939xee}, which classifies $\ln B_{10}$ as negligible 
($|\ln B_{10}| < 1$), positive to substantial ($1$--$3$), or strong to 
decisive ($|\ln B_{10}| > 5$). These thresholds serve as operational 
benchmarks throughout this work, with the understanding that 
$p(\mathcal{H}_{\mathrm{SM}}|D)$ is the quantity ultimately reported.

This construction also resolves a question that arises in any strategy 
scanning a large hypothesis space: the look-elsewhere effect (LEE). Within 
the Bayesian framework the correction is automatic: the evidence 
$p(D \mid \mathcal{H}_{\mathrm{NP}})$ is a weighted average over all SMEFT 
hypotheses, so a model that fits the data well locally contributes to $B_{10}$ 
only in proportion to its prior weight and to the degree that it outperforms 
the full ensemble of alternatives. A spurious excess in a single operator 
subset cannot drive a large Bayes factor unless it survives marginalisation 
over the entire hypothesis space. This is not a limitation of the framework 
but its correct behaviour: the Bayes factor reflects global statistical 
evidence for NP rather than the local fit quality of any individual operator 
subset.

\subsection{Correlation structure of the model posterior}
\label{sec:correlation}

The marginal inclusion probabilities $p(b_i = 1 \mid D)$ characterise the
support for individual operators, but one of the distinctive advantages of
working with a full posterior over model space is that it grants access to
richer information than any single-model analysis can provide. In
particular, $p(\vec{b} \mid D)$ encodes the relational structure of the
posterior: the degree to which the presence of one operator in a
high-posterior model is associated with the presence or absence of another.
This structure carries physically meaningful information about both the
sensitivity of the dataset and the operator patterns expected from UV
completions, information that is simply not accessible in traditional
global fits, where the operator content is fixed. Its systematic extraction from $p(\vec{b} \mid D)$ therefore 
represents one of the concrete advantages of the present framework 
over traditional analyses.

The natural starting point is the joint inclusion probability,
\begin{equation}
p(b_i = 1,\, b_j = 1 \mid D)
=
\sum_{\vec{b}:\, b_i = 1,\, b_j = 1} p(\vec{b} \mid D),
\end{equation}
which measures the posterior weight assigned to models in which both
$\mathcal{O}_i$ and $\mathcal{O}_j$ are simultaneously active. From this
object, two complementary diagnostics can be constructed, each probing a
different aspect of the operator correlation structure.

The first is the conditional inclusion probability,
\begin{equation}
p(b_j = 1 \mid b_i = 1, D)
=
\frac{p(b_i = 1,\, b_j = 1 \mid D)}{p(b_i = 1 \mid D)},
\end{equation}
which quantifies the posterior probability that $\mathcal{O}_j$ is active
given that $\mathcal{O}_i$ is. This is an asymmetric, directed quantity,
most informative when one operator is clearly favoured by the data and one
wishes to assess what its presence implies for related operators. In the
context of UV-motivated searches, a large conditional probability
$p(b_j = 1 \mid b_i = 1, D)$ could provide an early statistical indication that
$\mathcal{O}_j$ may also be present, even before the data are sufficiently
constraining to establish its marginal inclusion probability independently.

The second is the posterior correlation coefficient,
\begin{equation}
\phi_{ij}
=
\frac{
p(b_i = 1,\, b_j = 1 \mid D)
-
p(b_i = 1 \mid D)\, p(b_j = 1 \mid D)
}{
\sqrt{
p(b_i = 1 \mid D)\bigl(1 - p(b_i = 1 \mid D)\bigr)
\cdot
p(b_j = 1 \mid D)\bigl(1 - p(b_j = 1 \mid D)\bigr)
}
},
\end{equation}
the Pearson correlation coefficient of the binary inclusion variables $b_i$
and $b_j$ under the model posterior. By construction $\phi_{ij} \in [-1,1]$
and is symmetric in $i$ and $j$. Its sign and magnitude carry direct physical
meaning. A positive value indicates that $\mathcal{O}_i$ and $\mathcal{O}_j$ tend to appear together in high-posterior models more often than expected under independence. This is a statement about the joint sensitivity of the available observables: the data prefer both operators to be simultaneously active, irrespective of whether this reflects a correlation at the level of the UV completion, which $\phi_{ij}$ alone cannot establish. 
A negative value, by contrast,
signals that the two operators tend to be mutually exclusive in the
posterior: including one is sufficient to absorb the relevant data feature,
rendering the other redundant. Negative correlations are therefore a direct
signature of flat directions in operator space, helping in identifying precisely the
measurements whose inclusion would be most valuable for improving
operator-level resolution.

Together, these two quantities provide complementary handles on the
posterior landscape. The conditional inclusion probability is the natural
diagnostic when a specific operator has been identified as favoured and one
wishes to trace its implications for related operators. The posterior
correlation matrix $\phi_{ij}$
provides a global summary of which operators reinforce or compete with one
another across the full SMEFT basis.

\subsection{The matching scale as a parameter of inference}
\label{subsec:mu0}

Within the SMEFT framework, Wilson coefficients are defined at a reference scale $\mu_0$
from which the RGE is used to evolve their values to the energy scales probed by each
observable.
This scale is not fixed by the EFT construction: any choice of $\mu_0$ is formally
valid, and different choices are related by RGE running.
As argued in Sect.~\ref{subsec:smeft-hyp}, the natural choice in a UV-motivated analysis
is a high scale, reflecting the expectation that the effective operators are generated by
heavy dynamics at or above the TeV scale.
In practice, however, this still leaves open a continuous family of choices, since
the precise value of $\mu_0$ is not known a priori.

This ambiguity is not merely a technical nuisance.
Because the RGE induces operator mixing, the pattern of Wilson coefficients at low
energies depends on $\mu_0$: a coefficient that is generated at the matching scale will
spread into other directions of operator space as it is evolved downward, and the degree
of mixing is sensitive to the ratio of scales between $\mu_0$ and the relevant
experimental energies.
As a result, the statistical preference for particular operator subsets, and therefore
the output of model selection, carries an implicit dependence on the chosen reference
scale.

This observation suggests a natural extension of the model selection framework developed
in Ref.~\cite{Hirsch:2025qya} and in this work.
Rather than fixing $\mu_0$ by convention, one may treat it as a continuous parameter
of inference alongside the Wilson coefficients and the discrete operator content, and
allow the data to determine its favoured values.
It is important, however, to state precisely what such a posterior
on $\mu_0$ would represent. The reference scale is a parameter internal to the NP branch
of the hypothesis space: it labels where the active Wilson coefficients are defined, and
has no counterpart under $\mathcal{H}_{\mathrm{SM}}$, which contains no operators to run.
Any inference on $\mu_0$ is therefore necessarily conditional,
$p(\mu_0 \mid D, \mathcal{H}_{\mathrm{NP}})$,
and its relevance as a statement about the physical scale of new dynamics is gated by
the posterior weight of $\mathcal{H}_{\mathrm{NP}}$ itself. A sharply peaked
$p(\mu_0 \mid D, \mathcal{H}_{\mathrm{NP}})$ obtained while
$p(\mathcal{H}_{\mathrm{NP}} \mid D)$ remains small indicates only how the disfavoured
NP hypothesis would prefer to arrange itself were it true, not a property of the
mechanism generating the data. Only once a genuine signal has driven
$p(\mathcal{H}_{\mathrm{NP}} \mid D)$ to a non-negligible value does a peaked $\mu_0$
posterior become a meaningful, low-energy-only statement about $M_{\mathrm{NP}}$.


A further condition applies once $\mathcal{H}_{\mathrm{NP}}$ is itself supported by
the data: the preferred $\mu_0$ must track $M_{\mathrm{NP}}$ rather than merely
compensating boundary values chosen at the wrong scale. Absent a UV prior on the
Wilson coefficients, there is in principle considerable freedom to reproduce a given
pattern of deviations at an incorrect matching scale by adjusting the active
coefficients accordingly. We
argue, however, that the sparsity condition built into the model-selection framework
of Sect.~\ref{sec:models} provides a mechanism that
disfavours this freedom. RGE mixing generically disperses a sparse pattern of Wilson coefficients generated 
at the true scale $\mu_0 = M_{\mathrm{NP}}$ into a less sparse pattern when evolved 
to another reference scale. Consequently, reproducing the same low-energy deviations 
from an incorrect $\mu_0$ generally requires activating a broader combination of 
operators at that scale, rather than the original sparse subset. Because the model posterior explicitly penalises 
model complexity, this loss of sparsity disfavours incorrect choices of $\mu_0$ 
relative to the true scale, provided the dataset is sufficiently rich to resolve the 
resulting operator pattern. Establishing the quantitative reach of this mechanism, and its dependence on the
assumed sparsity pattern, operator basis, and EFT truncation, is
a necessary step before $\mu_0$ can be treated as a reliable proxy for $M_{\mathrm{NP}}$,
and is left for dedicated signal-injection studies in future work.


In the present work, we assess the sensitivity of our results to this choice 
by performing the analysis at two reference scales, $\mu_0 = 1\,\mathrm{TeV}$ 
and $\mu_0 = 10\,\mathrm{TeV}$, treating the resulting variation in
model posteriors as a diagnostic of where the assumed sparse SMEFT construction
most economically accommodates the data, conditional on $\mathcal{H}_{\mathrm{NP}}$.
The promotion of $\mu_0$ 
to a fully continuous parameter of inference, marginalised within the Bayesian 
evidence, remains
a natural and principled extension of this approach, to be deployed once a genuine
signal is established, and 
is left for future work.


\section{Methodology}
\label{sec:methodology}

The statistical framework developed in Sect.~\ref{sec:stat} identifies the 
Bayesian evidence $p(D \mid \vec{b})$ and the posterior over model space 
$p(\vec{b} \mid D)$ as the central objects of inference, but leaves open the 
question of how these quantities are computed in practice. We begin by 
describing the dataset and the EFT parametrisation that define the physical 
input to the analysis, including the operator basis, the treatment of linear 
and quadratic SMEFT contributions, and the role of renormalisation group 
evolution in connecting the matching scale to experimentally accessible 
energies. We then turn to the computational machinery. Exact evaluation of 
Bayesian evidences requires marginalisation over the continuous Wilson 
coefficient space for each of the $2^d$ candidate models, a computation that 
is doubly intractable: the per-model integrals are expensive, and the number 
of models is astronomically large. We address both bottlenecks through a 
sequence of well-motivated approximations: information criteria provide 
tractable proxies for the per-model evidence; a genetic algorithm replaces 
exhaustive enumeration of model space with a guided search that concentrates 
evaluations in regions of high posterior support; and a Hessian approximation 
replaces full posterior sampling within each model with an efficient Gaussian 
estimate. Together, these approximations make the framework of 
Sect.~\ref{sec:stat} computationally operational at the scale of $d = 52$ 
operators and several hundred data points, without sacrificing its essential 
statistical character.

\subsection{Dataset and EFT parametrisation}
\label{sec:dataset}

The analysis is carried out within the \textsc{SMEFiT} framework~\cite{Giani:2023gfq},
which provides a consistent implementation of dimension-six operator effects across a
broad range of collider observables. The dataset comprises $N_{\mathrm{dat}} = 417$ 
data points~\cite{terHoeve:2025gey}, including electroweak precision observables (EWPOs) from LEP\footnote{With respect to the analysis in Ref.~\cite{terHoeve:2025gey}, the theoretical predictions for EWPOs have been improved with updated values of the input parameters as well as the inclusion of parametric uncertainties and their correlations.} and Higgs, 
top-quark, and diboson measurements from LHC Run~2. The total covariance matrix 
$\Sigma$ accounts for statistical and systematic experimental uncertainties as well 
as theoretical uncertainties from missing higher-order SM corrections and parton 
distribution function errors.

For each candidate model $\vec{b}$, the agreement between theory predictions 
and data is quantified by the $\chi^2$ statistic,
\begin{equation}
\chi^2
=
\bigl(\mathbf{D} - \mathbf{T}_{\mathrm{EFT}}(\vec{c}_{\vec{b}})\bigr)^T
\Sigma^{-1}
\bigl(\mathbf{D} - \mathbf{T}_{\mathrm{EFT}}(\vec{c}_{\vec{b}})\bigr),
\end{equation}
where $\mathbf{D}$ is the vector of experimental measurements and 
$\mathbf{T}_{\mathrm{EFT}}(\vec{c}_{\vec{b}})$ the corresponding SMEFT 
predictions for the active operator content of $\vec{b}$.

We consider a basis of $d = 52$ CP-even dimension-six operators in the Warsaw
basis~\cite{Grzadkowski:2010es}, subject to a
$\mathrm{U}(2)_q \times \mathrm{U}(3)_d \times \mathrm{U}(2)_u \times
[\mathrm{U}(1)_l \times \mathrm{U}(1)_e]^3$ flavour symmetry assumption
(see Appendix~\ref{app:op-def} for explicit definitions).
Wilson coefficients are defined at the reference scale $\mu_0=10$ TeV for most of the analysis.
In Sec.~\ref{sec:mu0_results} we assess the effects of a different choice of $\mu_0$.

The analysis is performed at both linear and quadratic order in the Wilson
coefficients. At linear order, theory predictions take the form
\begin{equation}
\mathbf{T}_{\mathrm{EFT}}
=
\mathbf{T}_{\mathrm{SM}}
+
\sum_{i}
\frac{c_i(\mu_0)}{\Lambda^2}\,\mathbf{T}^{\mathrm{int}}_i,
\end{equation}
where $\mathbf{T}^{\mathrm{int}}_i$ encodes the SM--operator interference contribution.
At quadratic order, additional terms of the form
$c_i c_j \,\mathbf{T}^{\mathrm{quad}}_{ij} / \Lambda^4$ are included.
These squared contributions can help break degeneracies among operators that enter
similarly at linear level, and they extend the validity of the truncated EFT expansion
into parameter directions where the linear contribution is accidentally suppressed.
The linear-only analysis is reported alongside the quadratic results to isolate
the effect of the higher-order contributions.

Renormalisation group evolution (RGE) effects are systematically included at one-loop
order\footnote{Recently, the complete two-loop anomalous dimension matrix for dimension-six SMEFT operators has been computed~\cite{Born:2026xkr}, and its phenomenological implications have been studied in Refs.~\cite{Mantani:2026fao, Born:2026tgm}.}, connecting the high-energy matching scale where SMEFT operators arise from UV
completions to the lower-energy scales probed experimentally.
In practice, Wilson coefficients are evolved from $\mu_0$ to the relevant observable
scale for each data point using the Matrix Evolution
Approximation~\cite{terHoeve:2025gey}, which linearises the one-loop RGE flow and
provides an efficient implementation compatible with the large number of independent
fits required by the genetic algorithm search.

RGE induces operator mixing, allowing observables to receive contributions from
operators not directly generated at the matching scale~\cite{Battaglia:2021nys,Aoude:2022aro,DiNoi:2023onw,Garosi:2023yxg,Bartocci:2024fmm,Allwicher:2023shc,Maltoni:2024dpn,Greljo:2023bdy,Duhr:2025zqw}.
This is particularly relevant for electroweak precision observables at the $Z$-pole,
where the anticipated sub-permille precision of future facilities renders such
loop-induced effects a sensitive probe of new physics~\cite{Allwicher:2024sso}.
An important practical consequence is that operator mixing tends to suppress the
multimodal structure that commonly arises in the posterior of quadratic fits:
operators with large Wilson coefficients that could otherwise populate secondary
minima far from the SM tend to generate loop-induced contributions to precision
observables that are disfavoured by data, thereby removing these
modes~\cite{terHoeve:2025gey}.
This effect also improves the reliability of the Hessian approximation
described below.

\subsection{Information criteria for model comparison}
\label{sec:ic}

Exact evaluation of Bayesian evidences requires high-dimensional integrations and can be computationally demanding in large SMEFT model spaces.
For this reason, it is useful to consider approximate model comparison criteria that can be computed from maximum-likelihood fits, as discussed in Ref.~\cite{Hirsch:2025qya}.

The Akaike Information Criterion (AIC)~\cite{Akaike1974AIC} is defined as
\begin{equation}
\mathrm{AIC}
=
-2 \log \hat{\mathcal{L}} + 2k,
\end{equation}
where $\hat{\mathcal{L}}$ is the maximum likelihood and $k$ is the number of free parameters.
AIC is rooted in information theory and estimates the expected predictive performance of a model on new data.
Although it does not approximate the Bayesian evidence, differences in AIC can be used to define Akaike weights,
\begin{equation}
w_i
=
\frac{\exp(-\tfrac{1}{2}\Delta_i)}
     {\sum_j \exp(-\tfrac{1}{2}\Delta_j)},
\qquad
\Delta_i = \mathrm{AIC}_i - \min_j \mathrm{AIC}_j,
\end{equation}
which may be interpreted as relative model weights for predictive model averaging.

The Bayesian Information Criterion (BIC)~\cite{Schwarz:1978tpv} is given by
\begin{equation}
\mathrm{BIC}
=
-2 \log \hat{\mathcal{L}} + k \log N,
\end{equation}
where $N$ is the effective number of data points.
Under standard regularity assumptions and in the large-sample limit, BIC provides an approximation to the log evidence,
\begin{equation}
\log p(D \mid \vec{b})
\simeq
-\tfrac{1}{2}\,\mathrm{BIC} + \text{const}.
\end{equation}
As a consequence, BIC differences approximate log Bayes factors, and assuming equal model priors one obtains approximate posterior model probabilities,
\begin{equation}
p(\vec{b}_i \mid D)
\;\approx\;
\frac{\exp(-\tfrac{1}{2}\Delta_i)}
     {\sum_j \exp(-\tfrac{1}{2}\Delta_j)},
\qquad
\Delta_i = \mathrm{BIC}_i - \min_j \mathrm{BIC}_j .
\end{equation}

It should be emphasised that both AIC and BIC are best understood as heuristics rather
than exact statistical quantities.
The BIC approximation to the log evidence relies on regularity conditions and a
large-sample limit that may not be fully satisfied in the present setting, where the
number of data points is moderate ($N \sim 400$--$700$) and, when quadratic operator
contributions are included, the likelihood is no longer Gaussian in the Wilson
coefficients.
Similarly, while Akaike weights carry a Bayesian-flavoured interpretation, AIC does
not approximate the marginal likelihood and the weights should not be treated as
genuine posterior model probabilities.
In what follows, we adopt the BIC as our model selection criterion: its closer
correspondence to the Bayesian evidence and its stronger complexity penalty make it
better suited to the discovery-oriented goals of this analysis, where controlling
false positives is paramount. A direct comparison between the two criteria in the
context of BSM closure tests is presented in Ref.~\cite{Hirsch:2025qya}.

\subsection{Genetic algorithm search strategy}
\label{sec:ga}

The SMEFT model space $\mathcal{M} = \{0,1\}^d$ grows exponentially with the
number of operators and is therefore intractable to explore exhaustively: in our analysis with
$d = 52$, it contains $2^{52} \approx 4.5 \times 10^{15}$ candidate models.
To efficiently navigate this discrete and high-dimensional space, we employ a
genetic algorithm (GA)~\cite{Holland:1975,goldberg89, gad2021pygadintuitivegeneticalgorithm} that concentrates evaluations in regions of high
posterior support, replacing exhaustive enumeration with a guided stochastic
search.

The GA operates directly on the binary model vectors $\vec{b}$ defined in
Sect.~\ref{sec:models}. For each candidate model, the associated Wilson
coefficients $\vec{c}_{\vec{b}}$ are determined by minimising $\hat{\chi}^2(\vec{b})$
as defined in Sect.~\ref{sec:dataset}, and model fitness is assessed via the
penalised information criteria of Sect.~\ref{sec:ic}. The GA thus performs a
discrete search over model structures, while parameter optimisation is handled
deterministically within each evaluated model, with no functional relations or
UV-motivated correlations among Wilson coefficients imposed at this stage.

Starting from an initial population that includes by construction the SM and all
single-operator hypotheses, the GA evolves successive generations through three
standard operations: tournament selection with size $K = 4$, which preferentially
retains models with lower penalised scores; uniform crossover, which exchanges
bits between two parent strings; and bit-flip mutation with probability
$p_{\mathrm{mut}} = 0.025$, which maintains exploration of previously unvisited
regions. The population consists of 4000 candidate models evolved over 300
generations, with 300 parents selected at each step. We have verified the
robustness of our conclusions under moderate variations of these hyperparameters.

The output of the GA is a ranked ensemble of visited models $\{\vec{b}\}$.
In principle, the posterior $p(\vec{b} \mid D)$ defined in Sect.~\ref{sec:models}
is a sum over all $2^d$ models; in practice, this sum is approximated by
restricting it to the GA-visited subset, which by construction is concentrated
in the high-posterior region of model space. Model probabilities, operator
inclusion probabilities, model-averaged inferences and correlations are all constructed from
this ensemble as described in Sect.~\ref{sec:stat}. This approximation is
well-motivated provided the GA has achieved sufficient coverage of the
relevant region, a condition we assess both by verifying the
reproducibility of the inclusion probabilities across independent runs
and by direct comparison against exact results at reduced operator
dimensionality (see Appendix~\ref{app:ga_convergence}).

\subsection{Parameter optimisation and Hessian approximation}
\label{sec:hessian}

The analysis is performed at both linear and quadratic order in the Wilson 
coefficients, and the two cases differ in how the parameter optimisation is 
carried out. At linear order, the $\chi^2$ is a quadratic function of 
$\vec{c}_{\vec{b}}$ and the minimum $\hat{\vec{c}}_{\vec{b}}$ is obtained 
analytically via a linear system, making each model evaluation computationally 
inexpensive. At quadratic order, the additional terms render the 
objective non-quadratic in the parameters, and a numerical minimisation is 
required. Here we employ a standard gradient-based optimiser initialised at the 
SM solution.

Uncertainties on the Wilson coefficients are estimated via the Hessian 
approximation,
\begin{equation}
\mathrm{Cov}(\vec{c}_{\vec{b}})
\simeq
\left[
\frac{1}{2}\frac{\partial^2 \chi^2}{\partial c_i \,\partial c_j}
\right]^{-1}_{\!\vec{c}\, =\, \hat{\vec{c}}_{\vec{b}}},
\end{equation}
evaluated at the best-fit point $\hat{\vec{c}}_{\vec{b}}$, yielding Gaussian 
confidence intervals for each active coefficient. At linear order, where the 
$\chi^2$ is exactly quadratic in the Wilson coefficients, this is not an 
approximation: the Hessian is constant and the posterior is exactly Gaussian. 
At quadratic order, this is genuinely an approximation, 
whose accuracy we have validated against full posterior sampling for a 
representative set of benchmark models. Full Bayesian posterior 
sampling, as implemented in the standard \textsc{SMEFiT} pipeline, is 
computationally prohibitive in the present context where the GA evaluates 
$\mathcal{O}(10^5)$ distinct operator subsets over the course of a single run. 
The Hessian provides a tractable and well-motivated alternative, requiring only 
a single minimisation per model. Limitations are most apparent for operators with weak 
individual constraints, where non-Gaussian tails may be present. As noted in 
Sect.~\ref{sec:dataset}, however, the regularising effect of RGE operator mixing 
tends to suppress the multimodal structure that would otherwise make such 
approximations less trustworthy.

\section{Bayesian EFT analysis of LHC and LEP data}
\label{sec:eft_results}

We now apply the framework developed in 
Sects.~\ref{sec:stat} and~\ref{sec:methodology} to the combined dataset 
described in Sect.~\ref{sec:dataset}, comprising electroweak precision 
observables from LEP together with Higgs, top-quark, and diboson measurements 
from LHC Run~2. The analysis is performed at both linear and quadratic order 
in the Wilson coefficients, and results at the two orders are presented in 
parallel throughout to isolate the effect of the quadratic contributions at 
each level of inference. We proceed from the coarsest to the finest resolution 
of the model posterior: we begin with the global question of whether the data 
provide evidence for any departure from the SM, then examine which operators 
accumulate posterior support and what the model-averaged inference on their 
coefficients implies, before turning to the relational structure of the 
posterior encoded in the operator correlation matrix, and finally assessing the 
sensitivity of all these conclusions to the choice of matching scale $\mu_0$.

\subsection{Global evidence for new physics}
\label{sec:global_evidence}

\begin{figure}
    \centering
    \includegraphics[width=0.9\linewidth]{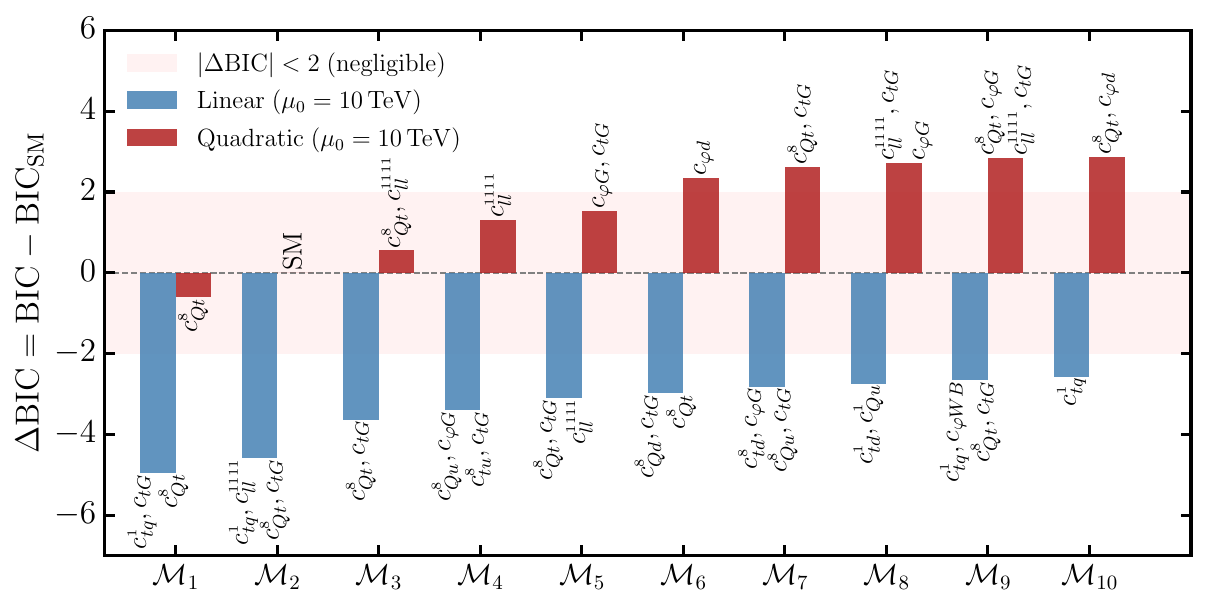}
    \caption{The ten top-ranked operator subsets visited by the GA, ordered by $\Delta\mathrm{BIC} = \mathrm{BIC}(\vec{b}) - \mathrm{BIC}_{\mathrm{SM}}$, for the linear (blue) and quadratic (red) fits at matching scale $\mu_0 = 10~\mathrm{TeV}$. Models with $\Delta\mathrm{BIC} < 0$ are preferred over the SM; the SM ($\Delta\mathrm{BIC} = 0$ by definition) is shown as a reference.}
    \label{fig:bayes_model}
\end{figure}

We begin by characterising the landscape of model posterior support without 
committing to any particular prior over the model space. Over the course of a single run, 
the GA visits around $2\times 10^5$ distinct operator subsets out of the $2^{52} \approx 
4.5\times 10^{15}$ models in the full space, concentrating evaluations in the 
high-posterior region as described in Sect.~\ref{sec:ga}. For each visited model 
$\vec{b}$, the difference
\begin{equation}
\Delta\mathrm{BIC}(\vec{b}) 
= 
\mathrm{BIC}(\vec{b}) - \mathrm{BIC}_{\mathrm{SM}}
\end{equation}
measures directly how much that operator subset improves the evidence relative 
to the SM, with no prior over models entering. Models with $\Delta\mathrm{BIC} < 0$ 
are preferred over the SM; those with $\Delta\mathrm{BIC} > 0$ are 
disfavoured. Fig.~\ref{fig:bayes_model} shows the ten top-ranked models ordered by $\Delta\mathrm{BIC}$ 
relative to the SM, for both the linear and quadratic fits, with $\mu_0 = 10~\mathrm{TeV}$.

In the linear fit, all ten top-ranked models outperform the SM, each improving the BIC by 
more than two units. The best-scoring model, comprising $c^1_{tq}$, $c^8_{Qt}$, and 
$c_{tG}$, achieves a BIC improvement of just under five units; close behind is a model 
that additionally includes $c^{1111}_{ll}$.
A clear structural pattern emerges across the top models: the favoured operator 
combinations are dominated by the colour-octet four-heavy quark operator $c^8_{Qt}$, 
the chromomagnetic operator $c_{tG}$, colour-octet and colour-singlet two-light-two-heavy 
four-fermion operators, and the four-lepton coefficient $c^{1111}_{ll}$. These 
results differ in two respects from the real-data application reported in 
Ref.~\cite{Hirsch:2025qya}. There, $c_{\varphi D}$ received some posterior support but 
this is no longer the case once the improved SM predictions for the EWPOs described 
in Sec.~\ref{sec:dataset} are adopted, which remove the mild tension previously 
responsible for its inclusion. Conversely, $c_{tG}$ now emerges as one of the leading operators. 
This follows from a sign error in its RGE evolution in the implementation used in 
Ref.~\cite{Hirsch:2025qya}, which has since been corrected.

Physically, 
the four-heavy operator improves the description of four-top and $t\bar{t}b\bar{b}$ production; the 
two-light-two-heavy operators and $c_{tG}$ improve the predictions for $t\bar{t}$ distributions; and $c^{1111}_{ll}$ reduces a residual tension in 
Bhabha scattering at LEP~\cite{ALEPH:2013dgf}. Notably, as shown in Sec.~\ref{sec:correlations}, $c_{tG}$ and $c^8_{Qt}$ are strongly 
correlated in the model posterior: neither operator achieves a sufficient improvement in 
isolation, and it is only their joint inclusion that yields the BIC gain needed to 
establish a clear preference over the SM. Furthermore, RGE evolution plays a central role in the improved description: without it, 
the preference for these operator combinations over the SM is significantly reduced. This is further discussed and analysed in Sec.~\ref{sec:mu0_results}
where the effects of the choice of $\mu_0$ are assessed.

The interpretation requires two caveats. First, RGE operator mixing causes four-fermion 
operators to feed into electroweak precision observables beyond tree level, so it is 
non-trivial that the Wilson coefficients driving these improvements do not simultaneously 
worsen the EWPO description. That these operators appear consistently across the top-ranked 
models indicates that the improvements are genuine and global: the relevant deviations are 
absorbed without introducing tensions elsewhere. Second, the favoured Wilson coefficients of the four-quark operators
are of order $1$--$10~\mathrm{TeV}^{-2}$, placing them squarely in the regime where 
quadratic contributions to the cross section are no longer negligible.

This is confirmed by the quadratic fit, where the picture changes substantially. Most 
operators that dominated the linear ranking all but disappear from the top models once the 
squared contributions are included, and the SM rises to become the second-best model 
overall. The only exception among the four-fermion operators is $c^{1111}_{ll}$, which 
remains among the top-ranked models: its Wilson coefficient was already in the linear 
regime, and the inclusion of quadratic corrections does not alter its ability to improve 
the description of the data. Beyond this, the single-operator model $\mathcal{O}^8_{Qt}$ 
marginally outperforms the SM by improving the description of four-top-quark production, 
though the preference remains well within the range of statistical fluctuations. Taken together, the linear and quadratic results tell a consistent story: the apparent 
preference for most four-fermion operators at linear order is an artefact of the truncation, 
driven by Wilson coefficients large enough that the neglected quadratic terms are in fact 
numerically significant. Once these are reinstated, the spurious improvement is removed 
and no operator subset clearly outperforms the SM. The posterior support is moreover 
distributed across several competing models rather than concentrated on a single dominant 
hypothesis, placing us precisely in the regime where BMA is most consequential: no single 
effective description is sufficiently favoured to be singled out, and inference must 
propagate the resulting model uncertainty.

To compress this landscape into a single discovery statistic, we follow the construction 
of Sect.~\ref{sec:discovery} and introduce a prior over the model space, assigning equal 
probability to the SM and to the composite NP hypothesis, with the latter distributed 
uniformly over the full set of models explored by the GA. The posterior probability of 
the SM then follows by weighting the evidence of each visited model against the chosen 
prior via Eq.~\eqref{eq:sm_posterior}. This separation between the prior-free landscape 
and the prior-dependent discovery statistic is one of the practical advantages of the 
framework, the GA-visited ensemble encodes all the information extracted from the data, 
and different prior choices amount simply to different ways of summarising it. At linear order we obtain 
$p(\mathcal{H}_{\mathrm{SM}}|D) = 99.5\%$ ($\ln B_{10} = -5.27$), and at quadratic 
order $p(\mathcal{H}_{\mathrm{SM}}|D) = 99.97\%$ ($\ln B_{10} = -8.10$), both 
corresponding to decisive evidence for the SM on the Jeffreys scale. These values confirm 
quantitatively what the $\Delta\mathrm{BIC}$ landscape already suggests qualitatively: 
the current dataset provides no statistically 
significant indication of any departure from the SM.

\subsection{Operator inclusion probabilities and BMA posteriors}
\label{sec:inclusion}

\begin{figure}
    \centering
    \vspace{-2cm}  
    \includegraphics[width=0.92\linewidth, page=1]{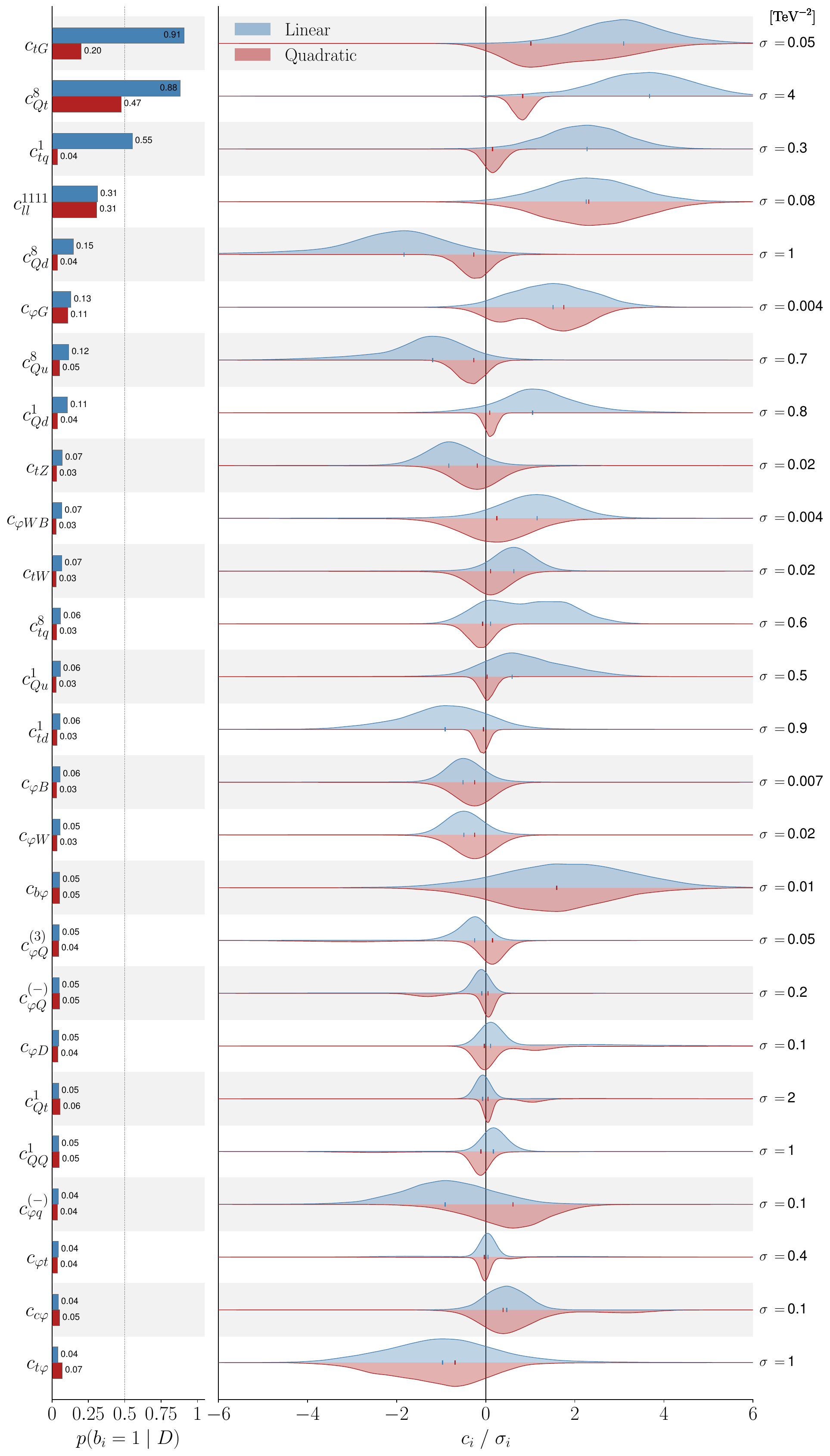}
    \caption{Marginal inclusion probabilities (left panel) and 
BMA conditional posteriors (right panel) for each 
operator, at linear (blue) and quadratic (red) EFT order, for $\mu_0 = 10\,\mathrm{TeV}$. 
Distributions are shown normalised to the standard deviation $\sigma_i$ 
of the linear conditional posterior.}
    \label{fig:inclusion1}
\end{figure}

\begin{figure}
    \centering
    \vspace{-2cm}  
    \includegraphics[width=0.92\linewidth, page=2]{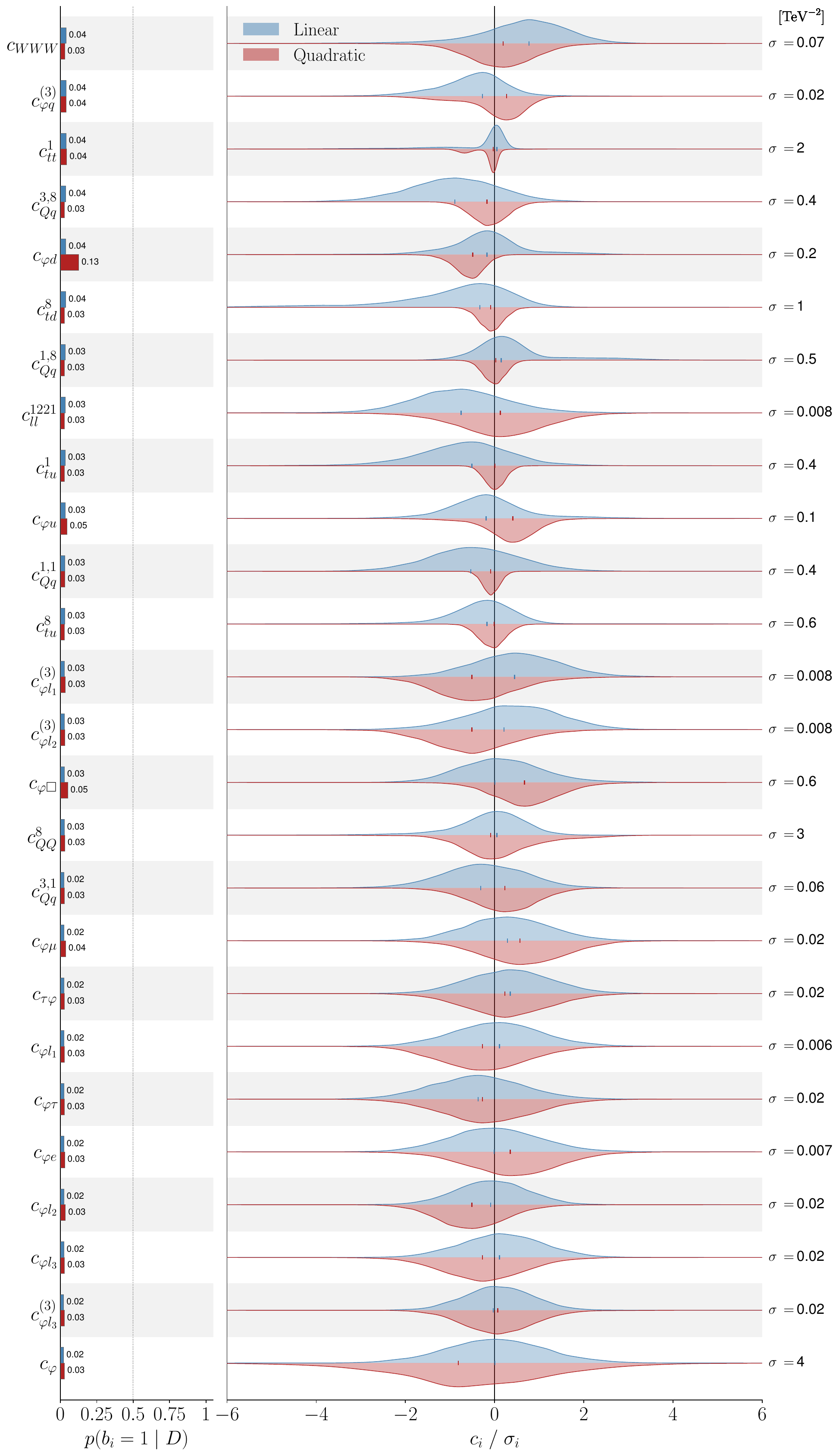}
    \caption{Same as Fig.~\ref{fig:inclusion1}}
    \label{fig:inclusion2}
\end{figure}

The marginal inclusion probability $p(b_i = 1 | D)$, defined in 
Sect.~\ref{sec:models}, provides the natural operator-level summary of the 
model posterior: it measures the degree to which the data support the 
presence of $\mathcal{O}_i$ after marginalising over all other operator 
combinations. Figs.~\ref{fig:inclusion1} and~\ref{fig:inclusion2}  display, for each operator, 
the inclusion probability alongside the conditional posterior 
$p(c_i | D, b_i = 1)$, the model averaged distribution of the Wilson coefficient 
given that the operator is present, both in the case of linear and quadratic fits. Wilson coefficients are displayed in order of marginal inclusion probability in the linear fit setup. All marginal distributions are shown normalised to the standard deviation of the linear 
posterior for ease of comparison. This two-layer presentation cleanly 
separates the question of operator existence from that of coefficient 
magnitude, as advocated in Sect.~\ref{sec:models}.

Several features of the inclusion probability spectrum deserve comment. At linear order, 
the operators with the highest marginal support are the chromomagnetic operator $c_{tG}$ 
and the colour-octet heavy-quark coefficient $c^8_{Qt}$, both reaching inclusion 
probabilities of approximately $90\%$, followed by $c^{1111}_{ll}$ and $c^1_{tq}$ at 
roughly $55\%$ and $30\%$ respectively. The remaining operators sit predominantly below 
the $10\%$ threshold, indicating that the data do not strongly prefer their presence once 
the complexity penalty is applied and marginalisation over all operator combinations is 
performed.

This structure is reflected in the conditional posteriors. Operators with 
higher inclusion probability display clear tension with the SM, with 
best-fit values displaced from zero by amounts consistent with the 
deviations suggested by the top models in Fig.~\ref{fig:bayes_model}. Operators with low 
inclusion probability have conditional posteriors broadly compatible with 
zero, as expected when the data provide no strong directional pull. Note that even at 
linear order, the BMA posteriors can be non-Gaussian and multimodal, since they 
are Gaussian mixtures weighted by model probabilities. This is most notably 
the case for $c_{tq}^8$, whose conditional posteriors 
exhibit a bimodal structure.

The inclusion probabilities at quadratic order follow a systematic pattern: almost all 
operators see their marginal support decrease when quadratic terms are included. This is 
expected, operators whose favoured Wilson coefficient values lie in the regime where 
squared contributions are non-negligible find that the large coefficients needed to 
improve the data description are no longer reachable once these terms are accounted for.
The notable exception is $c_{\varphi d}$, whose inclusion probability rises from $4\%$ 
to $13\%$, suggesting that this operator enters the data description more effectively 
once its quadratic contribution is available. Of particular interest is the reduction in the inclusion probability for $\mathcal{O}_{tG}$: 
while this operator is itself little affected by the inclusion of quadratic terms, 
it is strongly correlated with $c_{Qt}^8$ in the linear analysis, and the suppression 
of the latter at quadratic order propagates indirectly into a reduced marginal support for $\mathcal{O}_{tG}$.
The conditional posteriors $p(c_i \mid D, b_i = 1)$ characterise the implied Wilson 
coefficient values within each supported model, and tell a complementary story: a 
narrowing of the conditional posterior between linear and quadratic order is a direct 
signature of degeneracy-breaking, as the quadratic terms disfavour large-coefficient 
solutions that are consistent with the data at interference level alone.

\begin{figure}
    \centering
    \vspace{-2cm}  
    \includegraphics[width=0.92\linewidth, page=1]{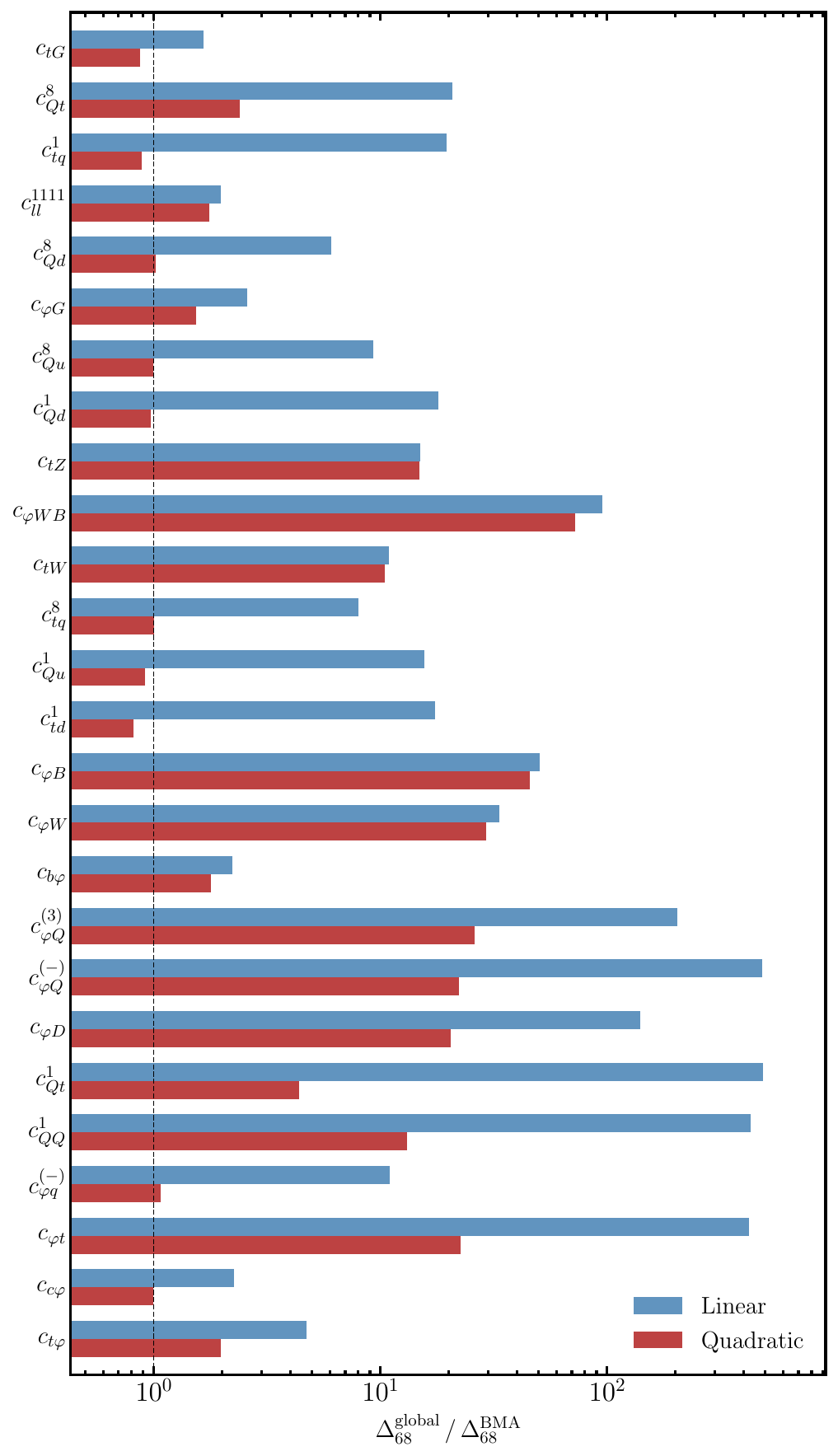}
    \caption{Ratio of the $68\%$ credible interval widths from BMA to those from a 
traditional global fit, for each operator at linear (blue) and quadratic (red) 
EFT order, with $\mu_0 = 10\,\mathrm{TeV}$.}
    \label{fig:comp_glob1}
\end{figure}

\begin{figure}
    \centering
    \vspace{-2cm}  
    \includegraphics[width=0.92\linewidth, page=2]{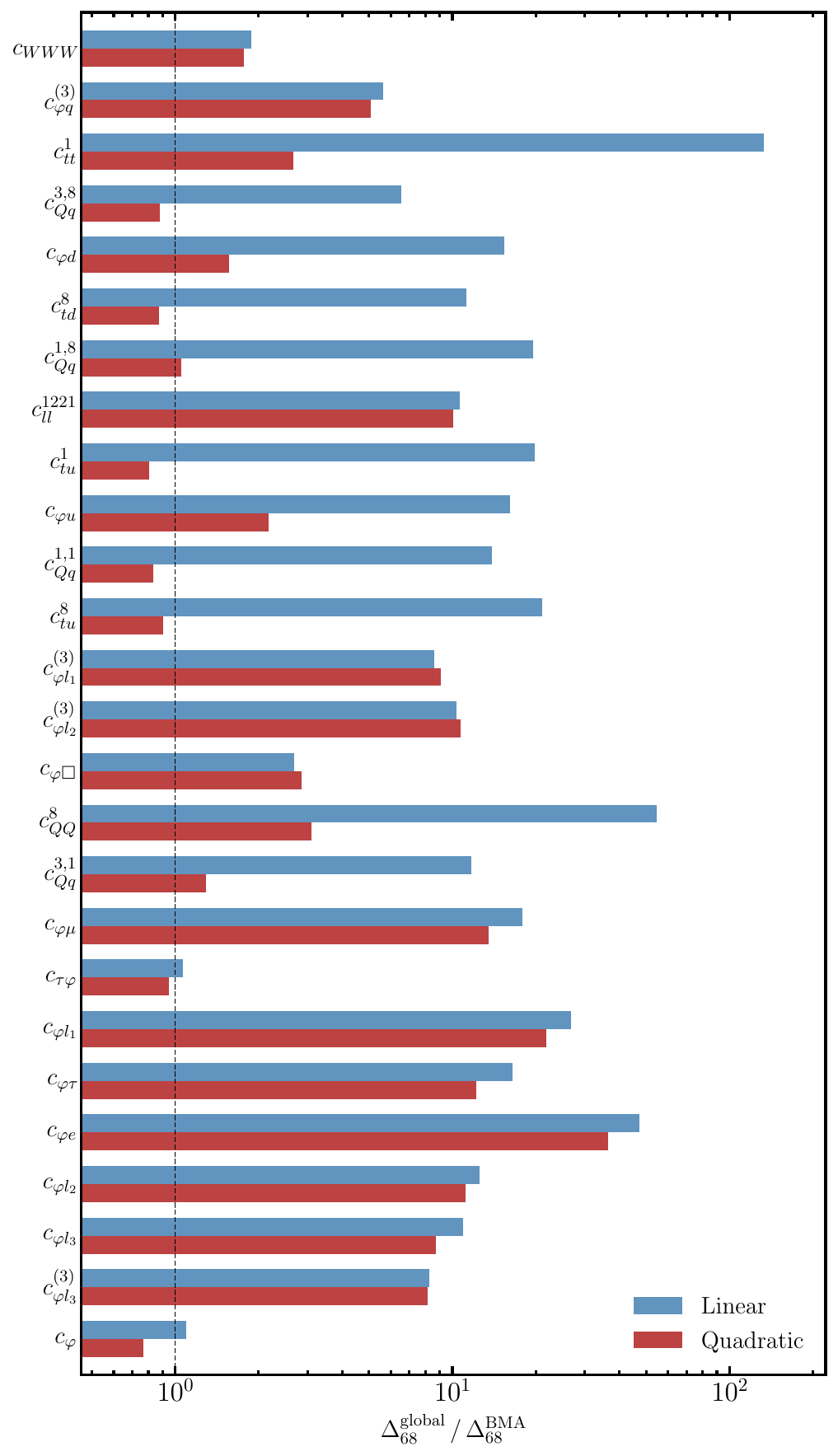}
    \caption{Same as Fig.~\ref{fig:comp_glob1}}
    \label{fig:comp_glob2}
\end{figure}

Figures~\ref{fig:comp_glob1} and~\ref{fig:comp_glob2} compare the width of the $68\%$ credible intervals obtained from BMA against those from a traditional global fit, expressed as their ratio. The comparison reveals a consistent and physically significant pattern. At linear order, BMA credible intervals are substantially narrower than their global-fit counterparts across the majority of operators, directly reflecting the improved characterisation potential of targeted model selection over full marginalisation: in the global fit, genuine NP contributions aligned with a narrow direction in operator space are diluted by the collective flexibility of the many weakly constrained directions, whereas BMA concentrates posterior weight on the operator combinations actually supported by the data.
At quadratic order, a subset of operators, predominantly two-light-two-heavy four-fermion interactions, exhibit the opposite behaviour, with BMA intervals modestly exceeding those of the global fit. This reversal is a distinctive feature of the quadratic regime: when operator-squared contributions are included, the dependence of the likelihood on the Wilson coefficients is no longer linear, and marginalisation over a non-Gaussian, potentially multimodal distribution can yield broader intervals than a global fit.
Crucially, the comparison is not limited to interval widths. In the global fit, not only are the bounds generally larger, but the central values of several coefficients are shifted relative to the BMA result, with tensions present in the BMA posteriors washed out entirely.

\subsection{Operator correlation structure}
\label{sec:correlations}

\begin{figure}
    \centering
    \vspace{-1cm}
    \includegraphics[width=0.75\linewidth]{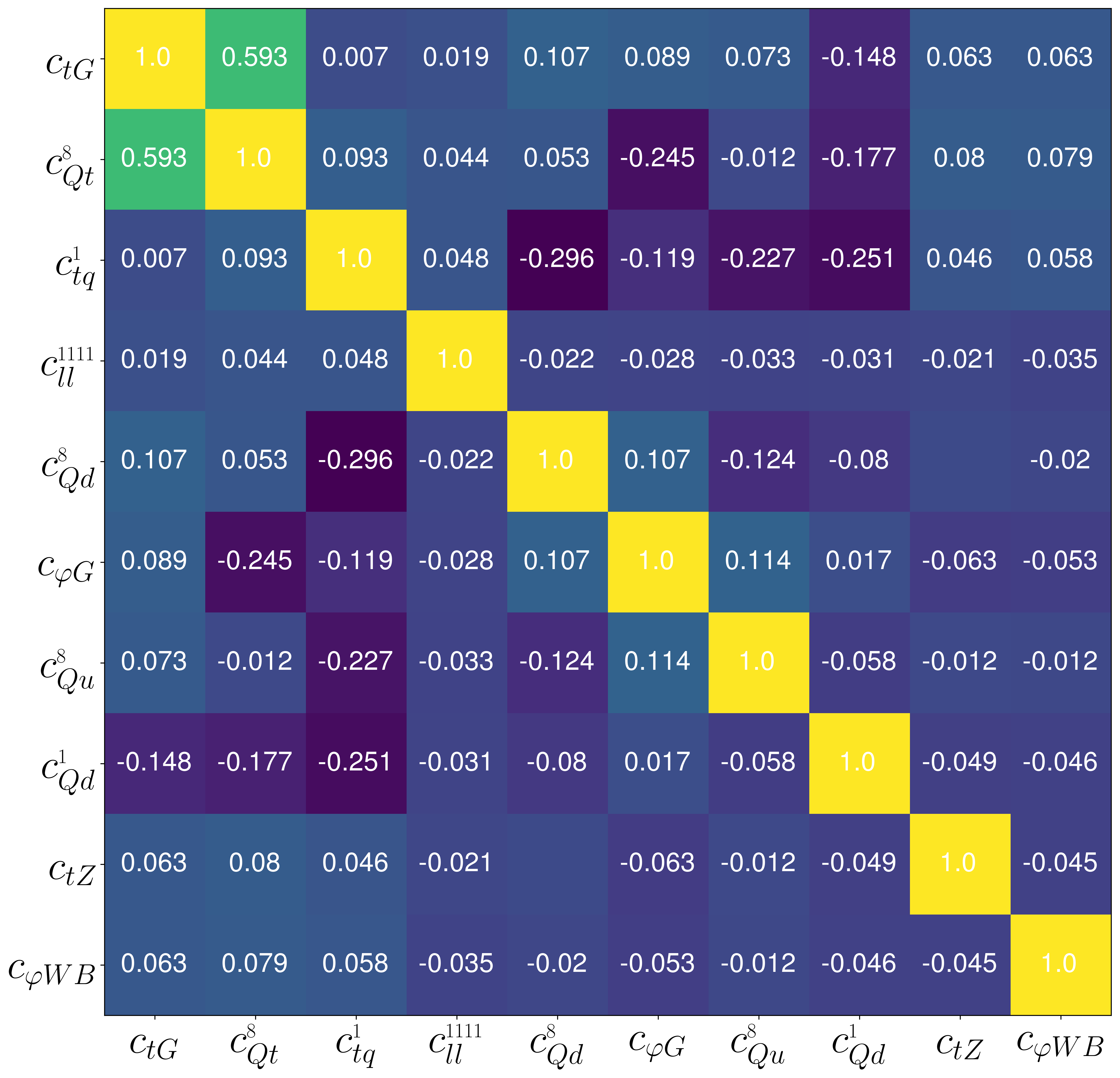}\\
    \vspace{0.5cm}
    \includegraphics[width=0.75\linewidth]{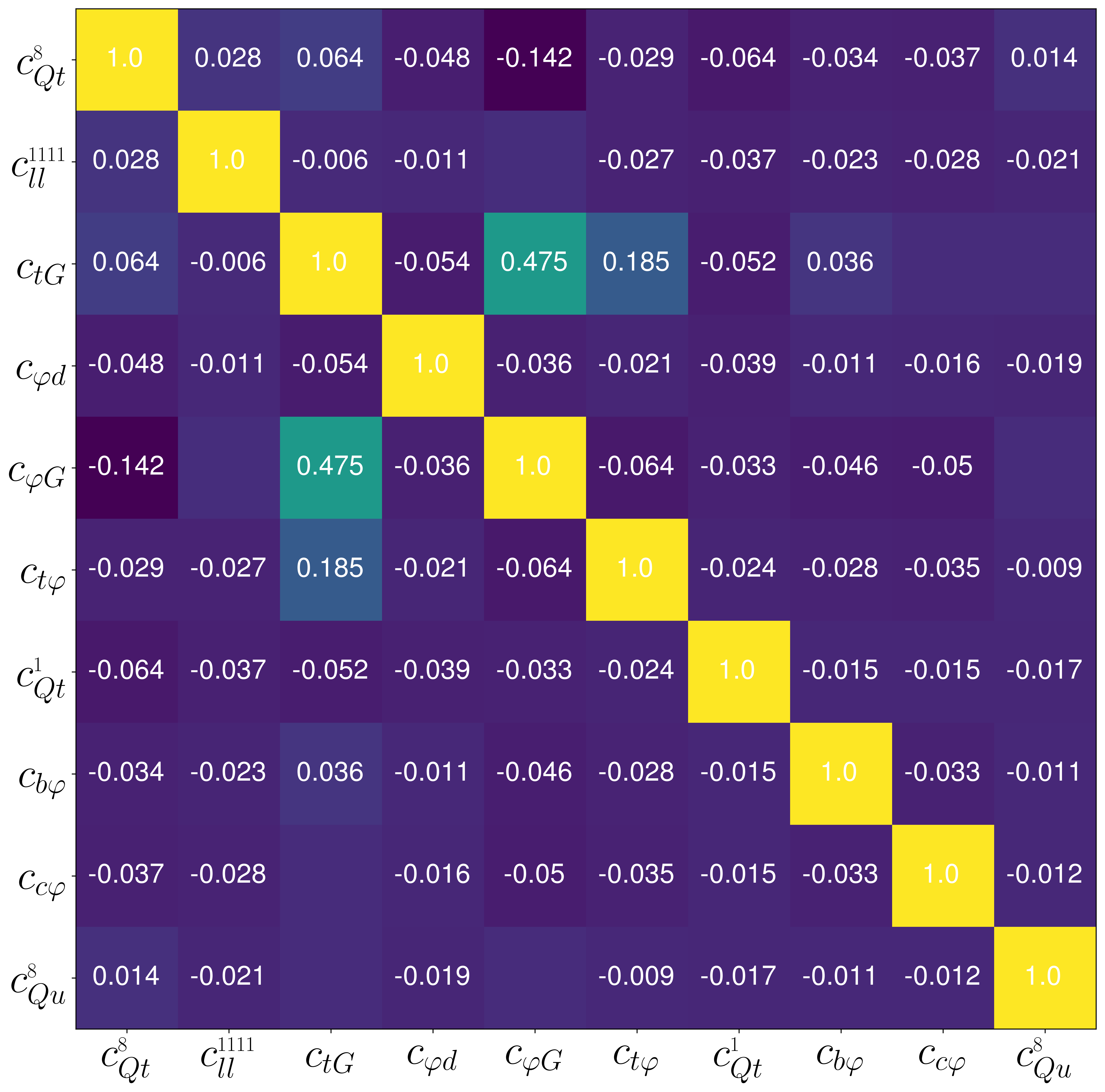}
    \caption{Posterior correlation matrix $\phi_{ij}$ for the ten operators with the 
highest marginal inclusion probabilities, at linear (top) and quadratic (bottom) 
EFT order. Positive entries indicate operator pairs that tend to co-appear in 
high-posterior models; negative entries signal mutually exclusive pairs that 
compete to absorb the same data feature.}
    \label{fig:correlation}
\end{figure}

The marginal inclusion probabilities of Sect.~\ref{sec:inclusion} 
characterise each operator individually, but one of the distinctive 
advantages of working with a full posterior over model space is access 
to the relational structure of the posterior. As discussed in 
Sect.~\ref{sec:correlation}, this is encoded in the posterior correlation 
matrix $\phi_{ij}$, which measures the degree to which operators tend to 
appear together or exclude one another across the high-posterior region 
of model space. Fig.~\ref{fig:correlation} displays $\phi_{ij}$ for the 
operators with the top 10 highest inclusion probabilities in the case of linear (top) and quadratic (bottom) fit.

Positive correlations identify operator pairs that tend to co-appear in high-posterior models, reflecting a joint sensitivity of the available observables rather than a property of the UV completion. A strong positive correlation may nonetheless carry indirect information about the underlying dynamics: if the data consistently prefer both operators together, this can help narrow the class of UV completions worth considering, even before either operator has accumulated individually decisive inclusion probability.
The correlation matrices reveal complementary structures at linear and quadratic order.
At linear order, the strongest positive correlation is observed between $c_{tG}$ and
$c_{Qt}^8$ ($\phi_{ij} \simeq 0.6$), while at quadratic order a positive correlation
of $\phi_{ij} \simeq 0.48$ emerges between $c_{tG}$ and $c_{\varphi G}$. In both
cases, activating $\mathcal{O}_{tG}$ induces, through one-loop mixing, a contribution
to $c_{\varphi G}$ that affects the description of Higgs data and must be
compensated by a partner operator, preventing $\mathcal{O}_{tG}$ from accumulating
strong inclusion probability in isolation. The compensating role is played by
different operators in each regime: at linear order it is $\mathcal{O}_{Qt}^8$,
whose own mixing into the top Yukawa operator $\mathcal{O}_{t\varphi}$ provides
the necessary rebalancing, while at quadratic order $\mathcal{O}_{\varphi G}$
enters directly at tree level in Higgs production.

Negative correlations carry a complementary message: they identify flat 
directions in operator space, where including one operator is sufficient 
to absorb the relevant data feature, rendering the other redundant. These 
are precisely the directions where additional measurements would be most 
valuable in improving operator-level resolution. A strong negative 
correlation between $\mathcal{O}_i$ and $\mathcal{O}_j$ implies that the 
current dataset cannot simultaneously determine whether both are present, 
and points directly to the type of observable, one that couples 
differently to the two operators, whose inclusion would lift the 
degeneracy. Among the operators receiving the highest inclusion probabilities, however, no pronounced negative correlations are observed: the largest in magnitude reaches only $\phi_{ij} \simeq -0.3$, between $c_{tq}^1$ and $c_{Qd}^8$ in the linear fit, suggesting that the top-ranked operator combinations are not significantly degenerate with one another under the current dataset.

The comparison between the linear and quadratic correlation matrices reveals the concrete impact of squared contributions on the operator degeneracy structure. At linear order, correlations are generically stronger, reflecting the extended flat directions that arise when two operators enter observables with similar interference patterns. At quadratic order, the squared and cross terms break many of these degeneracies by introducing curvature into directions of operator space that are flat at linear level: the correlation between $c_{tG}$ and $c_{Qt}^8$, for instance, is substantially reduced.

\subsection{Dependence on the matching scale}
\label{sec:mu0_results}

\begin{figure*}
    \centering
    \hspace*{-2cm}
    \includegraphics[width=0.54\textwidth, page=1]{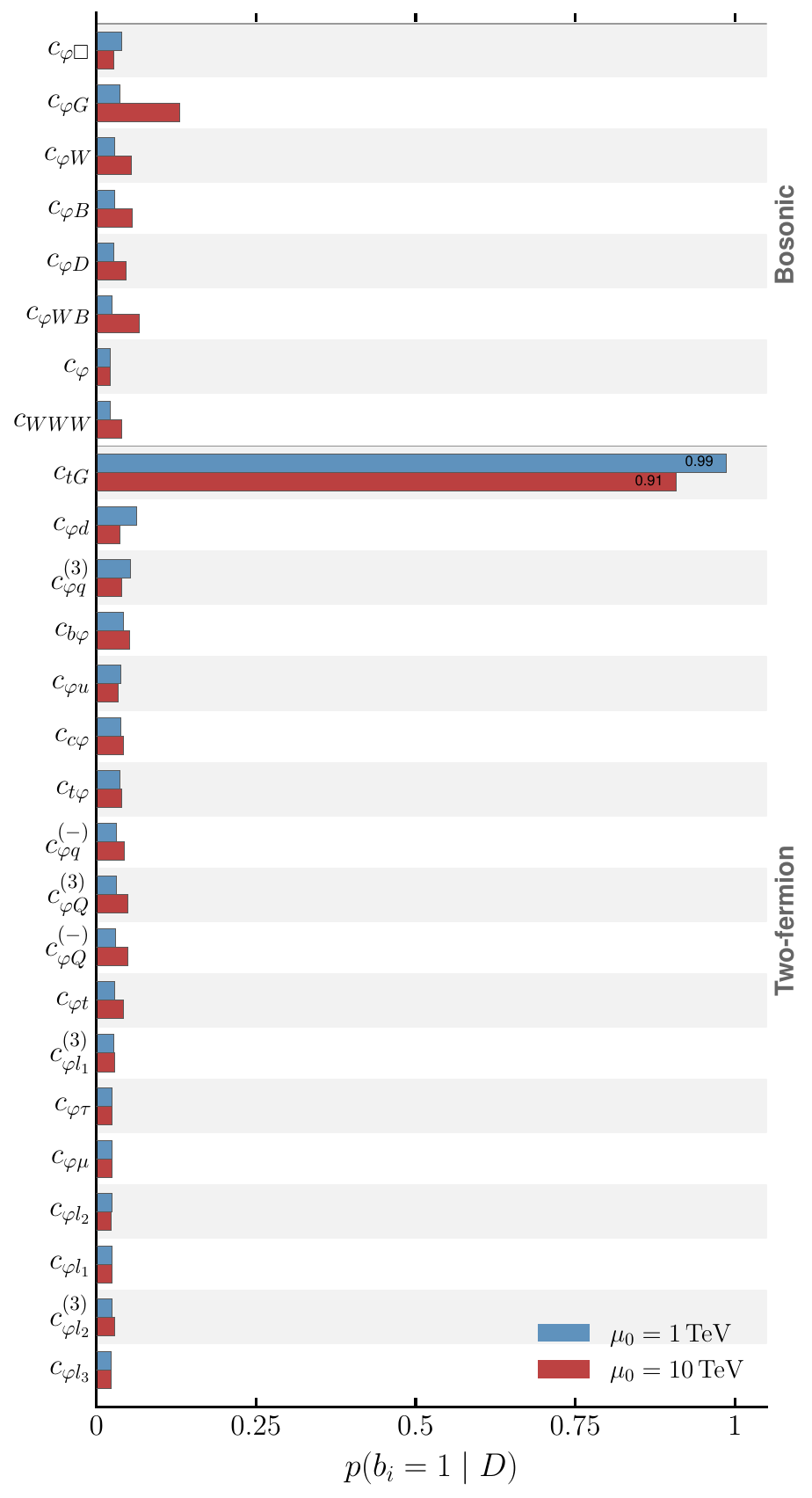}
    \hspace{0.5cm}
    \includegraphics[width=0.54\textwidth, page=2]{figures/inclusion_prob_mu0_comparison.pdf}
    \hspace*{-2cm}
    \caption{Operator inclusion probabilities $p(b_i = 1 \mid D)$ at linear order, 
    comparing the reference scales $\mu_0 = 1\,\mathrm{TeV}$ (blue) and 
    $\mu_0 = 10\,\mathrm{TeV}$ (red). Operators are grouped by class and ordered 
    within each group by decreasing inclusion probability at $\mu_0 = 1\,\mathrm{TeV}$.  Operators whose bars are stable 
    across the two scales have posterior support driven by direct, tree-level 
    sensitivity of the data; those with pronounced variation reflect the 
    sensitivity of operator mixing under RGE to the choice of matching scale.}
    \label{fig:mu0_dependence_lin}
\end{figure*}

\begin{figure*}
    \centering
    \hspace*{-2cm}
    \includegraphics[width=0.54\textwidth, page=3]{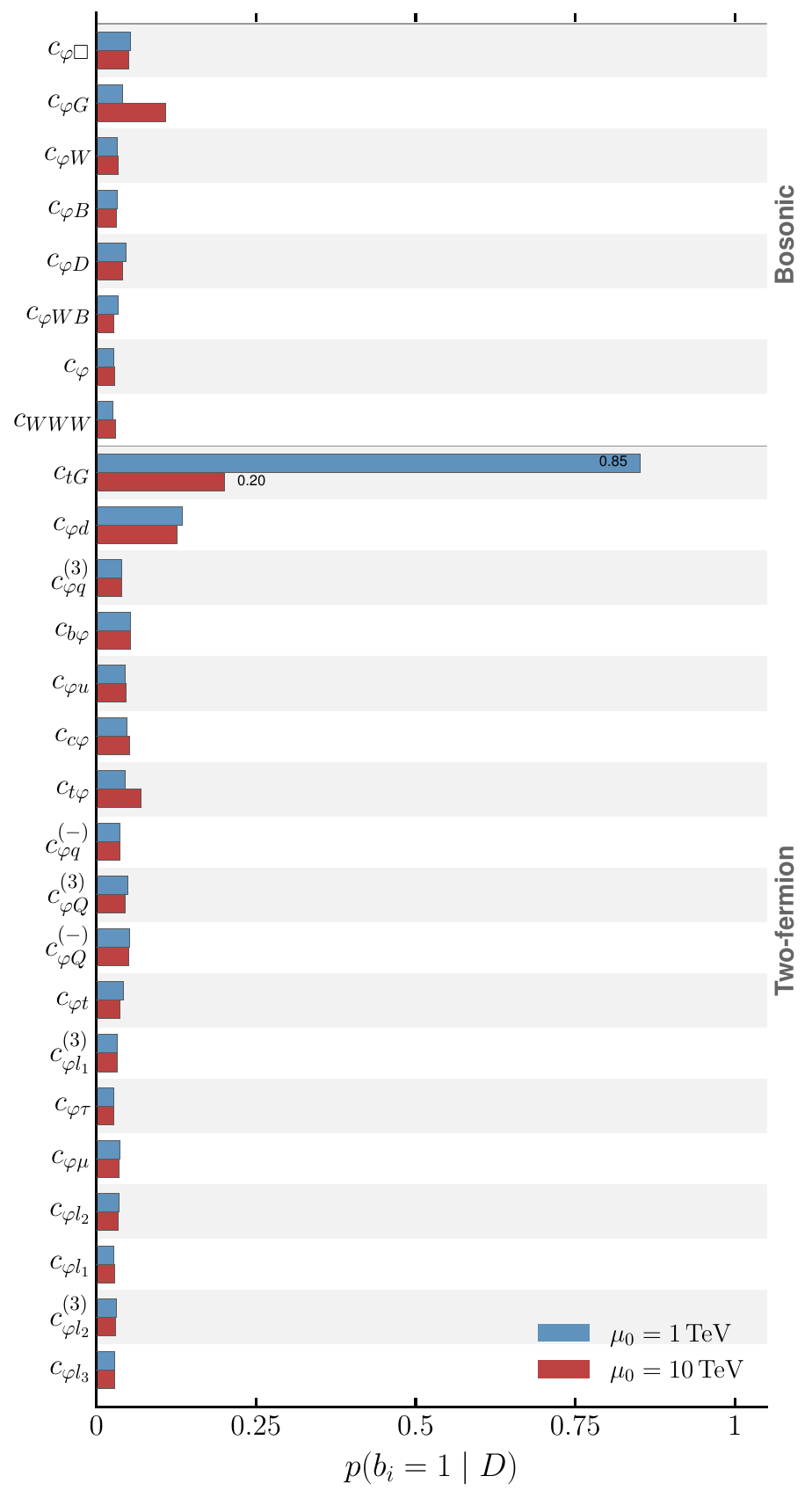}
    \hspace{0.5cm}
    \includegraphics[width=0.54\textwidth, page=4]{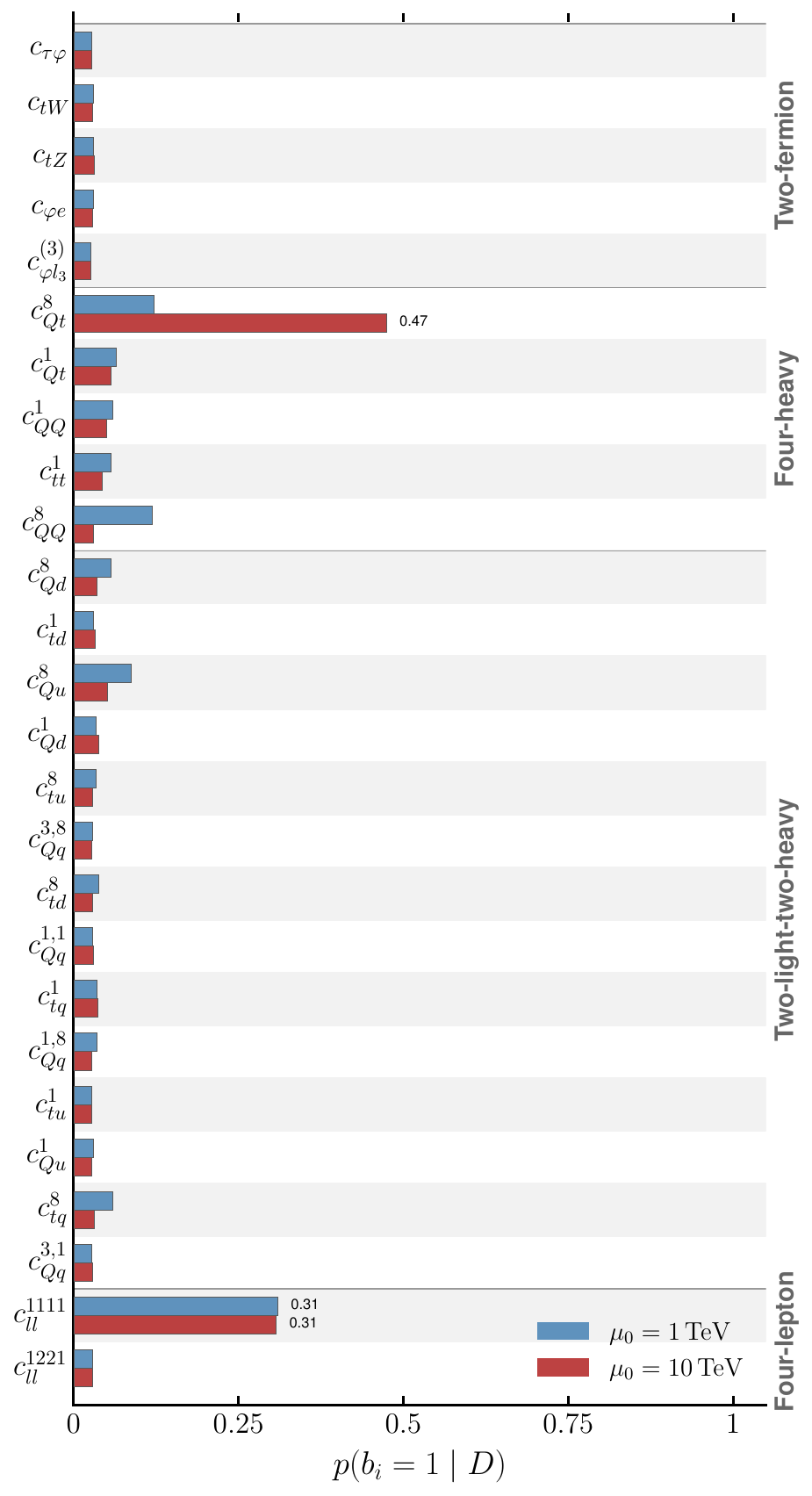}
    \hspace*{-2cm}
    \caption{Same as Fig.~\ref{fig:mu0_dependence_lin}, but for the quadratic EFT analysis. 
    The inclusion of squared operator contributions modifies the sensitivity to individual 
    operators and can resolve degeneracies present at linear order, leading in some cases 
    to qualitatively different $\mu_0$ dependence.}
    \label{fig:mu0_dependence_quad}
\end{figure*}

The results presented in the preceding subsections are obtained at a fixed reference
scale $\mu_0 = 10\,\mathrm{TeV}$. As discussed in Sect.~\ref{subsec:mu0}, this choice
is not fixed by the EFT construction: different values of $\mu_0$ induce different
patterns of operator mixing under RGE, which propagate into the statistical preference
for particular operator subsets. We now assess the sensitivity of our results to this
choice by repeating the analysis at $\mu_0 = 1\,\mathrm{TeV}$, treating the resulting
variation in model posteriors as a probe of the choice of the reference scale.

Figs.~\ref{fig:mu0_dependence_lin} and~\ref{fig:mu0_dependence_quad} display the
operator inclusion probabilities $p(b_i = 1 \mid D)$ at both scales, for the linear
and quadratic analyses respectively. Two qualitatively distinct patterns emerge.

A first group of operators exhibits inclusion probabilities that are stable across
the two values of $\mu_0$. This stability is physically significant: it indicates
that the posterior support for these operators is driven by a direct, tree-level
imprint on the data rather than by RGE-induced mixing into other directions, and
therefore constitutes evidence for their relevance independently of the specific $\mu_0$ scale. The four-lepton operator $c_{ll}^{1111}$ is
a clear example at both orders.

A second group shows a more pronounced $\mu_0$ dependence. This sensitivity is a
direct consequence of RGE mixing: varying $\mu_0$ changes how much a coefficient
at the matching scale spreads into other operator directions at experimental energies,
altering the effective sensitivity of the dataset to each operator and reshaping the
relative posterior weight assigned to competing hypotheses through marginalisation.
Whether this $\mu_0$ dependence can be interpreted as evidence for the
physical scale of new dynamics, as discussed in Sect.~\ref{subsec:mu0}, depends on
whether $\mathcal{H}_{\mathrm{NP}}$ itself is supported by the data, a question we
address below.

The chromomagnetic operator $c_{tG}$ illustrates how this behaviour can differ
qualitatively between the linear and quadratic analyses. At linear order, its
inclusion probability rises modestly from $91\%$ at $\mu_0 = 10\,\mathrm{TeV}$ to
$99\%$ at $\mu_0 = 1\,\mathrm{TeV}$, a shift consistent with a mild RGE-induced
enhancement. At quadratic
order, the same operator exhibits a markedly stronger dependence: its inclusion
probability falls from approximately $85\%$ at $\mu_0 = 1\,\mathrm{TeV}$ to around
$20\%$ at $\mu_0 = 10\,\mathrm{TeV}$. This is a direct consequence of the
$\mathcal{O}_{tG}$--$\mathcal{O}_{\varphi G}$ correlation identified in
Sect.~\ref{sec:correlation}: through one-loop mixing, $c_{tG}$ feeds into
$c_{\varphi G}$, and when running from $10\,\mathrm{TeV}$ the accumulated mixing
becomes large enough that a coefficient sized to resolve the $t\bar{t}$ tensions
simultaneously distorts the description of Higgs data, strongly suppressing the
inclusion of $\mathcal{O}_{tG}$ at the higher matching scale.

The four-quark sector provides a further illustration of $\mu_0$-driven
reshaping of the posterior. Four-quark operators enter the data at two distinct
levels: at tree level, they contribute directly to $t\bar{t}$ and four-top
production, while through RGE mixing they develop loop-induced contributions to
electroweak precision observables and, in the case of $\mathcal{O}_{Qt}^8$, to
Higgs production via mixing into the top Yukawa operator. The tension between
these two roles is controlled by $\mu_0$: a larger scale separation amplifies
the mixing contributions, making it increasingly difficult for a single operator
to simultaneously provide a good description of the observables it enters at tree
level and those it contaminates through running. This competition reshapes the
posterior across the four-quark sector as $\mu_0$ varies. At linear order,
$c_{Qt}^8$ receives substantial posterior support at $\mu_0 = 10\,\mathrm{TeV}$
but loses it at $\mu_0 = 1\,\mathrm{TeV}$, where support shifts instead to the
two-light-two-heavy operators, which are less affected by mixing-induced tension at the
lower scale. At quadratic order, $c_{Qt}^8$ similarly loses support at the lower scale, but
no compensating redistribution to two-light-two-heavy operators emerges. This
is because the alternative hypotheses that absorb the support at linear order
require large Wilson coefficients to fit the data, and at quadratic order such
large values are disfavoured by the squared contributions.

This pattern has a direct consequence for the global discovery statistic: the stronger 
preference for the SM at $\mu_0 = 10\,\mathrm{TeV}$ relative to $\mu_0 = 1\,\mathrm{TeV}$ 
is genuinely driven by RGE-induced mixing, which grows with the scale separation between 
$\mu_0$ and the experimental scales. Some operator subsets provide a good tree-level 
description of the mild deviations present in the data, but once evolved through a larger 
mixing window their induced contributions to other observables become incompatible with 
the data. Since the SM posterior is computed relative to the full composite NP hypothesis 
of Sect.~\ref{sec:discovery}, the resulting exclusion of these subsets at larger $\mu_0$ 
mechanically increases the posterior weight assigned to the SM.

The global discovery statistic $p(\mathcal{H}_{\mathrm{SM}} \mid D)$ and the
corresponding log Bayes factor $\ln B_{10}$ are reported as a function of $\mu_0$
in Table~\ref{tab:mu0_bayes}, for both EFT orders. The relative stability of
$p(\mathcal{H}_{\mathrm{SM}} \mid D)$ across scales indicates that the global
evidential picture is robust to the choice of matching scale. At linear order, we
find $p(\mathcal{H}_{\mathrm{SM}} \mid D) = 94.6\%$ ($\ln B_{10} = -2.86$) at
$\mu_0 = 1\,\mathrm{TeV}$, to be compared with $99.5\%$ at
$\mu_0 = 10\,\mathrm{TeV}$; at quadratic order, the corresponding values are
$99.8\%$ ($\ln B_{10} = -6.75$) and $99.97\%$. In both cases, the data show
slightly more tension with the SM at the lower matching scale, though the shift
is modest and the SM remains strongly supported at both scales. On the Jeffreys
scale, all four configurations fall in the range of positive-to-substantial
evidence for the SM, with no configuration approaching the threshold for
evidence against it. Since $\mathcal{H}_{\mathrm{NP}}$ is disfavoured at
both scales, this $\mu_0$ dependence should be read as identifying where the
assumed sparse construction most economically accommodates the residuals, not
as evidence for a physical NP scale. As noted in Sect.~\ref{subsec:mu0}, the promotion of $\mu_0$ 
to a fully continuous parameter of inference, marginalised within the Bayesian 
evidence, represents a natural and principled extension of this framework and is 
left for future work.

\begin{table}[h]
\centering
\renewcommand{\arraystretch}{1.4}
\begin{tabular}{llcc}
\hline
EFT order & $\mu_0$ & $p(\mathcal{H}_{\mathrm{SM}} \mid D)$ & $\ln B_{10}$ \\
\hline
\multirow{2}{*}{Linear}    
    & $1\,\mathrm{TeV}$  & $94.6\%$ & $-2.86$ \\
    & $10\,\mathrm{TeV}$ & $99.5\%$ & $-5.27$ \\[2pt]
\multirow{2}{*}{Quadratic} 
    & $1\,\mathrm{TeV}$  & $99.8\%$ & $-6.75$ \\
    & $10\,\mathrm{TeV}$ & $99.97\%$ & $-8.10$ \\
\hline
\end{tabular}
\caption{Global SM posterior probability $p(\mathcal{H}_{\mathrm{SM}} \mid D)$ 
and log Bayes factor $\ln B_{10}$ at two reference scales $\mu_0$, 
for both the linear and quadratic EFT analyses. Equal priors 
$p(\mathcal{H}_{\mathrm{SM}}) = p(\mathcal{H}_{\mathrm{NP}}) = 1/2$ are 
assumed throughout.}
\label{tab:mu0_bayes}
\end{table}

\section{Conclusions and outlook}
\label{sec:conclusions}

We have presented a comprehensive framework for indirect NP discovery in the SMEFT, 
grounded in Bayesian model comparison and implemented via a genetic algorithm search 
over a discrete space of $2^{52}$ operator subsets. The central thesis of the paper is 
that SMEFT analyses aimed at discovery should be understood not as parameter estimation 
within a fixed high-dimensional model, but as inference over a structured space of 
competing hypotheses, each corresponding to a physically distinct low-energy realisation 
of potential UV dynamics. The statistical tools developed here, operator inclusion 
probabilities, BMA posteriors, posterior correlation matrices, and the SM posterior 
probability as a discovery statistic, give concrete expression to this perspective.

Applied to the combined dataset of LEP electroweak precision observables and LHC 
Run~2 Higgs, top-quark, and diboson measurements, the framework yields a clear and 
consistent picture. At both linear and quadratic order in the Wilson coefficients, the 
global SM posterior probability $p(\mathcal{H}_{\mathrm{SM}} \mid D)$ places the 
current data in the regime of positive-to-decisive evidence for the SM on the Jeffreys 
scale, with no operator subset accumulating statistically compelling evidence for NP. 
The parallel linear and quadratic analysis proves essential for establishing this 
conclusion: the mild preference for four-fermion operators visible at linear order is 
unambiguously identified as an artefact of the EFT truncation, driven by Wilson 
coefficient values large enough that quadratic corrections are numerically significant. 
Once these are reinstated, the apparent improvement is removed and the SM is 
comfortably restored as the preferred description of the data.

The operator-level analysis reveals a richer structure beneath this global null result. 
At linear order, $\mathcal{O}_{tG}$ and $\mathcal{O}_{Qt}^8$ accumulate the highest 
marginal inclusion probabilities, reaching approximately $90\%$ each. Their strong 
posterior correlation, neither achieves a sufficient BIC improvement in isolation, is 
driven by RGE mixing: the one-loop evolution of $c_{tG}$ into $c_{\varphi G}$ forces 
a partner operator to compensate the resulting distortion of Higgs observables. At 
quadratic order, the squared contributions break these correlations, suppress large 
Wilson coefficients, and substantially reorganise the operator hierarchy. The exception 
is $c_{ll}^{1111}$, whose inclusion probability is stable across both EFT orders and 
the two analysed matching scales, marking it as the most robustly preferred operator in the current dataset. 
The global discovery statistic 
remains stable across scales, confirming that the null result is not an artefact 
of the matching scale choice.

The comparison with a traditional global fit underscores a key advantage of the present 
approach. BMA credible intervals are consistently narrower at linear order, 
concentrating posterior weight on the operator combinations actually supported by the 
data rather than diluting them through marginalisation over weakly constrained 
directions. Tensions visible in the BMA posteriors are washed out entirely in the 
global fit, underscoring that the two approaches are not interchangeable when the goal 
is discovery and characterisation rather than conservative bounding.

The promotion of the matching scale $\mu_0$ to a continuous parameter of 
inference, marginalised within the Bayesian evidence, would, once a
genuine signal for $\mathcal{H}_{\mathrm{NP}}$ is established, transform the scale 
dependence observed here into a direct statistical statement about the preferred 
scale of new dynamics, derived entirely from low-energy observables. A complementary direction concerns the basis dependence of model selection. 
As argued in Sect.~\ref{sec:epistemology}, what is genuinely basis-dependent 
is the granularity of the hypothesis space: working in a fixed basis favours 
hypotheses aligned with its on/off switches, and physically motivated directions 
that cut across basis operators may be less efficiently explored. This opens 
the prospect of identifying operator bases that are optimally adapted to 
discovery, concentrating posterior weight most efficiently on the relevant 
directions in hypothesis space.

One concrete instance of such a basis is a BSM-particle basis, constructed from 
the field content of a candidate UV completion and matched onto SMEFT via the 
corresponding dictionaries~\cite{deBlas:2017xtg, Guedes:2023azv, Guedes:2024vuf}. 
This optimality is itself UV-dependent, however: it holds when the new dynamics 
is well approximated by a small number of weakly coupled resonances, but need not 
hold otherwise, for instance for strongly coupled completions with no such 
particle content to match onto. A BSM-particle basis is therefore best regarded 
not as a replacement for a SMEFT-operator basis, but as a complementary strategy, 
well adapted to discovery when the weakly coupled assumption holds and poorly 
adapted otherwise. The Warsaw basis adopted throughout this work remains a sensible
default in this broader context: it is a standard, complete, non-redundant
convention with mature matching and RGE machinery, presupposing no specific
realisation of the new dynamics. Its operators are nonetheless representatives of
equivalence classes under field redefinitions rather than physical observables, so
sparsity in this basis carries no basis-independent meaning. More broadly, no single hypothesis space, whether organised around explicit 
particle content, an operator basis, or some other UV-motivated parametrisation, 
offers a privileged measure of simplicity: each is suited to a different class 
of UV assumptions.

On the 
experimental side, the framework is well positioned to assess the discovery reach of 
the HL-LHC and future lepton colliders: as datasets grow in precision and diversity, 
the operator correlation structure identified here will sharpen, degeneracies will be 
lifted, and the model posterior will concentrate, providing a quantitative and evolving 
picture of indirect discovery potential.

\section*{Acknowledgements}
We thank S. Tentori, M. Hirsch, F. Maltoni, V. Sanz and E. Vryonidou for useful discussions.
L.M. acknowledges support from the European Union under the MSCA fellowship (Grant agreement N. 101149078) {\it Advancing global SMEFT fits in the LHC precision era (EFT4ward)}.

\newpage
\appendix
\section{SMEFT operator definition}
\label{app:op-def}

The operator basis adopted in this analysis consists of CP-even dimension-six operators in the Warsaw basis~\cite{Grzadkowski:2010es}, subject to a $\text{U}(2)_q \times \text{U}(3)_d \times \text{U}(2)_u \times (\text{U}(1)_\ell \times \text{U}(1)_e)^3$ flavour symmetry assumption, as implemented in the \texttt{SMEFiT} framework~\cite{Giani:2023gfq}. The complete set of operators and their explicit definitions is collected in the tables below: purely bosonic operators in Table~\ref{tab:oper_bos}, four-fermion operators in Table~\ref{tab:oper_fourtop}, two-fermion and four-lepton operators in Tables~\ref{tab:oper_ferm_bos} and~\ref{tab:oper_ferm_bos2}.

\begin{table}[H] 
  \begin{center}
  \footnotesize
    \renewcommand{\arraystretch}{1.3}
        \begin{tabular}{lll|lll}
          \toprule
          Operator & Coefficient & Definition & Operator & Coefficient & Definition \\
        \midrule\midrule
        $\Op{\varphi G}$ & $c_{\varphi G}$  & $\left( \varphi^\dagger \varphi - \frac{v^2}{2} \right)G^{\mu\nu}_{\sss A}\,G_{\mu\nu}^{\sss A}$ 
        & 
        $\Op{\varphi \square}$ & $c_{\varphi \square}$ & $(\pdp)\square(\pdp)$ \\
        $\Op{\varphi B}$ & $c_{\varphi B}$ & $\left( \varphi^\dagger \varphi - \frac{v^2}{2} \right) B^{\mu\nu}\,B_{\mu\nu}$
        &
        $\Op{\varphi D}$ & $c_{\varphi D}$ & $(\varphi^\dagger D^\mu\varphi)^\dagger(\varphi^\dagger D_\mu\varphi)$ \\ 
        $\Op{\varphi W}$ &$c_{\varphi W}$ & $\left( \varphi^\dagger \varphi - \frac{v^2}{2} \right)W^{\mu\nu}_{\sss I}\,W_{\mu\nu}^{\sss I}$ 
        &
        $\mathcal{O}_{W}$& $c_{WWW}$ & $\epsilon_{IJK}W_{\mu\nu}^I W^{J,\nu\rho} W^{K,\mu}_\rho$ \\ 
        $\Op{\varphi W B}$ &$c_{\varphi W B}$ & $(\varphi^\dagger \tau_{\sss I}\varphi)\,B^{\mu\nu}W_{\mu\nu}^{\sss I}$
        &
        $\Op{\varphi}$ & $c_{\varphi}$ & $\left( \varphi^\dagger \varphi - \frac{v^2}{2} \right)^3$ \\
       \bottomrule
        \end{tabular}
        \caption{Bosonic operators at dimension six that modify Higgs dynamics and electroweak gauge interactions.
          \label{tab:oper_bos}}
\end{center}
\end{table}

\begin{table}[H] 
  \begin{center}
  \footnotesize
    \renewcommand{\arraystretch}{1.35}
        \begin{tabular}{ll|ll}
          \toprule
          DoF & Definition (Warsaw basis notation) & DoF & Definition (Warsaw basis notation) \\
          \midrule\midrule
      $c_{QQ}^1$    & $2\ccc{1}{qq}{3333}-\frac{2}{3}\ccc{3}{qq}{3333}$ 
      &
      $c_{QQ}^8$    & $8\ccc{3}{qq}{3333}$\\  
      $c_{Qt}^1$    & $\ccc{1}{qu}{3333}$
      &
      $c_{Qt}^8$    & $\ccc{8}{qu}{3333}$\\   
      \midrule      
      $c_{Qq}^{1,8}$ & $\ccc{1}{qq}{i33i}+3\ccc{3}{qq}{i33i}$  
      &
      $c_{Qq}^{1,1}$ & $\ccc{1}{qq}{ii33}+\frac{1}{6}\ccc{1}{qq}{i33i}+\frac{1}{2}\ccc{3}{qq}{i33i}$\\    
      $c_{Qq}^{3,8}$ & $\ccc{1}{qq}{i33i}-\ccc{3}{qq}{i33i}$  
      &
      $c_{Qq}^{3,1}$ & $\ccc{3}{qq}{ii33}+\frac{1}{6}(\ccc{1}{qq}{i33i}-\ccc{3}{qq}{i33i})$\\     
      $c_{tq}^{8}$   & $\ccc{8}{qu}{ii33}$ 
      &
      $c_{tq}^{1}$   & $\ccc{1}{qu}{ii33}$\\   
      $c_{tu}^{8}$   & $2\ccc{}{uu}{i33i}$ 
      &
      $c_{tu}^{1}$   & $\ccc{}{uu}{ii33}+\frac{1}{3}\ccc{}{uu}{i33i}$\\   
      $c_{Qu}^{8}$   & $\ccc{8}{qu}{33ii}$
      &
      $c_{Qu}^{1}$   & $\ccc{1}{qu}{33ii}$\\    
      $c_{td}^{8}$   & $\ccc{8}{ud}{33jj}$ 
      &
      $c_{td}^{1}$   & $\ccc{1}{ud}{33jj}$\\    
      $c_{Qd}^{8}$   & $\ccc{8}{qd}{33jj}$ 
      &
      $c_{Qd}^{1}$   & $\ccc{1}{qd}{33jj}$\\
         \bottomrule
  \end{tabular}
  \caption{Definitions of the four-fermion coefficients entering the fit. The coefficients are grouped into two categories: four-heavy operators (upper part) and two-light–two-heavy operators (lower part). The flavour indices $i$ and $j$ take values $i = 1, 2$ and $j = 1, 2, 3$, respectively.
\label{tab:oper_fourtop}}
  \end{center}
\end{table}

\begin{table}[H]
  \begin{center}
  \footnotesize
    \renewcommand{\arraystretch}{1.3}
    \begin{tabular}{p{1.4cm} p{1.3cm} p{4.2cm} | p{1.4cm} p{1.3cm} p{4.2cm}}
      \toprule
      Operator & Coefficient & \quad Definition & Operator & Coefficient & \quad Definition \\
      \midrule\midrule
      \multicolumn{6}{c}{3rd generation quarks} \\
      \midrule\midrule
    $\Op{\varphi Q}^{(1)}$ & $c_{\varphi Q}^{(1)}$~(*) & $i\big(\varphi^\dagger\lra{D}_\mu\,\varphi\big)\big(\bar{Q}\,\gamma^\mu\,Q\big)$ 
    &
    $\Op{tW}$ & $c_{tW}$ & $i\big(\bar{Q}\tau^{\mu\nu}\,\tau_{\sss I}\,t\big)\,\tilde{\varphi}\,W^I_{\mu\nu}+ \text{h.c.}$ \\ 
    $\Op{\varphi Q}^{(3)}$ & $c_{\varphi Q}^{(3)}$  & $i\big(\varphi^\dagger\lra{D}_\mu\,\tau_{\sss I}\varphi\big)\big(\bar{Q}\,\gamma^\mu\,\tau^{\sss I}Q\big)$ 
    &
    $\Op{tB}$ & $c_{tB}$~(*) & $i\big(\bar{Q}\tau^{\mu\nu}\,t\big)\,\tilde{\varphi}\,B_{\mu\nu}+ \text{h.c.}$\\ 
    $\Op{\varphi t}$ & $c_{\varphi t}$ & $i\big(\varphi^\dagger\,\lra{D}_\mu\,\varphi\big)\big(\bar{t}\,\gamma^\mu\,t\big)$
    &
    $\Op{t G}$ & $c_{tG}$ & $ig{\sss S}\,\big(\bar{Q}\tau^{\mu\nu}\,T_{\sss A}\,t\big)\,\tilde{\varphi}\,G^A_{\mu\nu}+ \text{h.c.}$ \\ 
    $\Op{t \varphi}$ & $c_{t\varphi}$ & $\left(\pdp\right)\bar{Q}\,t\,\tilde{\varphi} + \text{h.c.}$ 
    &
    $\Op{b \varphi}$ & $c_{b\varphi}$ & $\left(\pdp\right)\bar{Q}\,b\,\varphi + \text{h.c.}$ \\  
    \midrule\midrule
    \multicolumn{6}{c}{1st, 2nd generation quarks} \\
    \midrule\midrule
    $\Op{\varphi q}^{(1)}$ & $c_{\varphi q}^{(1)}$~(*) & $\sum\limits_{\sss i=1,2} i\big(\varphi^\dagger\lra{D}_\mu\,\varphi\big)\big(\bar{q}_i\,\gamma^\mu\,q_i\big)$ 
    &
    ${\Op{\varphi d}}$ & ${{c_{\varphi d}}}$ & $\sum\limits_{\sss i=1,2,3} i\big(\varphi^\dagger\,\lra{D}_\mu\,\varphi\big)\big(\bar{d}_i\,\gamma^\mu\,d_i\big)$\\ 
    $\Op{\varphi q}^{(3)}$ & $c_{\varphi q}^{(3)}$ & $\sum\limits_{\sss i=1,2} i\big(\varphi^\dagger\lra{D}_\mu\,\tau_{\sss I}\varphi\big)\big(\bar{q}_i\,\gamma^\mu\,\tau^{\sss I}q_i\big)$
    &
    $\Op{c \varphi}$ & $c_{c \varphi}$ & $\left(\pdp\right)\bar{q}_2\,c\,\tilde\varphi + \text{h.c.}$ \\ 
    ${\Op{\varphi u}}$ & ${{c_{\varphi u}}}$ & $\sum\limits_{\sss i=1,2} i\big(\varphi^\dagger\,\lra{D}_\mu\,\varphi\big)\big(\bar{u}_i\,\gamma^\mu\,u_i\big)$\\ 
    \midrule\midrule
    \multicolumn{6}{c}{two-leptons} \\
    \midrule\midrule
    $\Op{\varphi \ell_i}$ & $c_{\varphi \ell_i}$ & $i\big(\varphi^\dagger\lra{D}_\mu\,\varphi\big)\big(\bar{\ell}_i\,\gamma^\mu\,\ell_i\big)$ 
    &
    $\Op{\varphi \mu}$ & $c_{\varphi \mu}$ & $i\big(\varphi^\dagger\lra{D}_\mu\,\varphi\big)\big(\bar{\mu}\,\gamma^\mu\,\mu\big)$ \\  
    $\Op{\varphi \ell_i}^{(3)}$ & $c_{\varphi \ell_i}^{(3)}$ & $i\big(\varphi^\dagger\lra{D}_\mu\,\tau_{\sss I}\varphi\big)\big(\bar{\ell}_i\,\gamma^\mu\,\tau^{\sss I}\ell_i\big)$ 
    &
    $\Op{\varphi \tau}$ & $c_{\varphi \tau}$ & $i\big(\varphi^\dagger\lra{D}_\mu\,\varphi\big)\big(\bar{\tau}\,\gamma^\mu\,\tau\big)$ \\  
    $\Op{\varphi e}$ & $c_{\varphi e}$ & $i\big(\varphi^\dagger\lra{D}_\mu\,\varphi\big)\big(\bar{e}\,\gamma^\mu\,e\big)$ 
    &
    $\Op{\tau \varphi}$ & $c_{\tau \varphi}$ & $\left(\pdp\right)\bar{\ell_3}\,\tau\,{\varphi} + \text{h.c.}$ \\
    \midrule\midrule
    \multicolumn{6}{c}{four-leptons} \\
    \midrule\midrule
    $\Op{\ell\ell}$ & $c_{\ell\ell}$ & $\left(\bar \ell_1\gamma_\mu \ell_2\right) \left(\bar \ell_2\gamma^\mu \ell_1\right)$ 
    &
    $\Op{\ell\ell}^{11}$ & $c^{11}_{\ell\ell}$ & $\left(\bar \ell_1\gamma_\mu \ell_1\right) \left(\bar \ell_1 \gamma^\mu \ell_1\right)$ \\
    \bottomrule
\end{tabular}
\caption{Same as Table~\ref{tab:oper_bos}, but for operators containing two fermion fields, either quarks or leptons, as well as the four-lepton operator $\mathcal{O}_{\ell\ell}$ and $\mathcal{O}^{11}_{\ell\ell}$. The flavour index $i$ runs from 1 to 3. Coefficients marked with an asterisk (*) do not correspond to independent physical degrees of freedom in the fit, but are instead replaced by $c_{\varphi q}^{(-)}$, $c_{\varphi Q}^{(-)}$, and $c_{tZ}$, as defined in Table~\ref{tab:oper_ferm_bos2}.
\label{tab:oper_ferm_bos}}
\end{center}
\end{table}

\begin{table}[H]
  \begin{center}
  \footnotesize
    \renewcommand{\arraystretch}{1.3}
    \begin{tabular}{l l}
    \toprule
     Coefficient \quad & Definition\\ 
     \midrule\midrule
    $c_{\varphi Q}^{(-)}$ & $c_{\varphi Q}^{(1)}-c_{\varphi Q}^{(3)}$\\ 
    $c_{tZ}$ & $-s_\theta \, c_{tB}+ c_\theta \,c_{tW}$ \\ 
    $c_{\varphi q}^{(-)}$ & $c_{\varphi q}^{(1)}-c_{\varphi q}^{(3)}$ \\ 
    \bottomrule
\end{tabular}
\caption{Linear combinations of the two-fermion operators listed in Table~\ref{tab:oper_ferm_bos}, which replace those marked with an asterisk (*) at the fit level.
\label{tab:oper_ferm_bos2}}
\end{center}
\end{table}
\section{Convergence of the genetic algorithm search}
\label{app:ga_convergence}

The GA search introduced in Sect.~\ref{sec:ga} is a stochastic procedure: the
sequence of populations visited over successive generations depends on the
random seed governing tournament selection, crossover, and mutation. Since
the model posterior $p(\vec{b} \mid D)$ underlying the results of
Sect.~\ref{sec:eft_results} is constructed from the ensemble of models
visited by a single run of the GA, it is necessary to verify that this
ensemble is representative of the high-posterior region of
$\{0,1\}^{52}$, rather than an artefact of a particular random
realisation of the search.

To this end, we repeat the GA search ten times on the LHC+LEP dataset at
linear EFT order and $\mu_0 = 10\,\mathrm{TeV}$, varying only the random
seed and keeping all other hyperparameters fixed at the values quoted in
Sect.~\ref{sec:ga}: a population of 4000 models evolved over 300
generations, with 300 parents selected at each step via tournament
selection ($K = 4$) and a mutation probability $p_{\mathrm{mut}} = 0.025$.
For each operator $\mathcal{O}_i$, we compute the marginal inclusion
probability $p(b_i = 1 \mid D)$ separately from each of the ten runs and
define the spread
\begin{equation}
\Delta p_i = \max_r\, p_r(b_i = 1 \mid D) \;-\; \min_r\, p_r(b_i = 1 \mid D),
\end{equation}
where $r = 1, \ldots, 10$ labels the independent runs. This quantity
directly measures the sensitivity of the reported inclusion probabilities
to the stochastic nature of the search, as distinct from their genuine
statistical uncertainty under the data.

\begin{figure}
    \centering
    \includegraphics[width=0.85\linewidth, page=1]{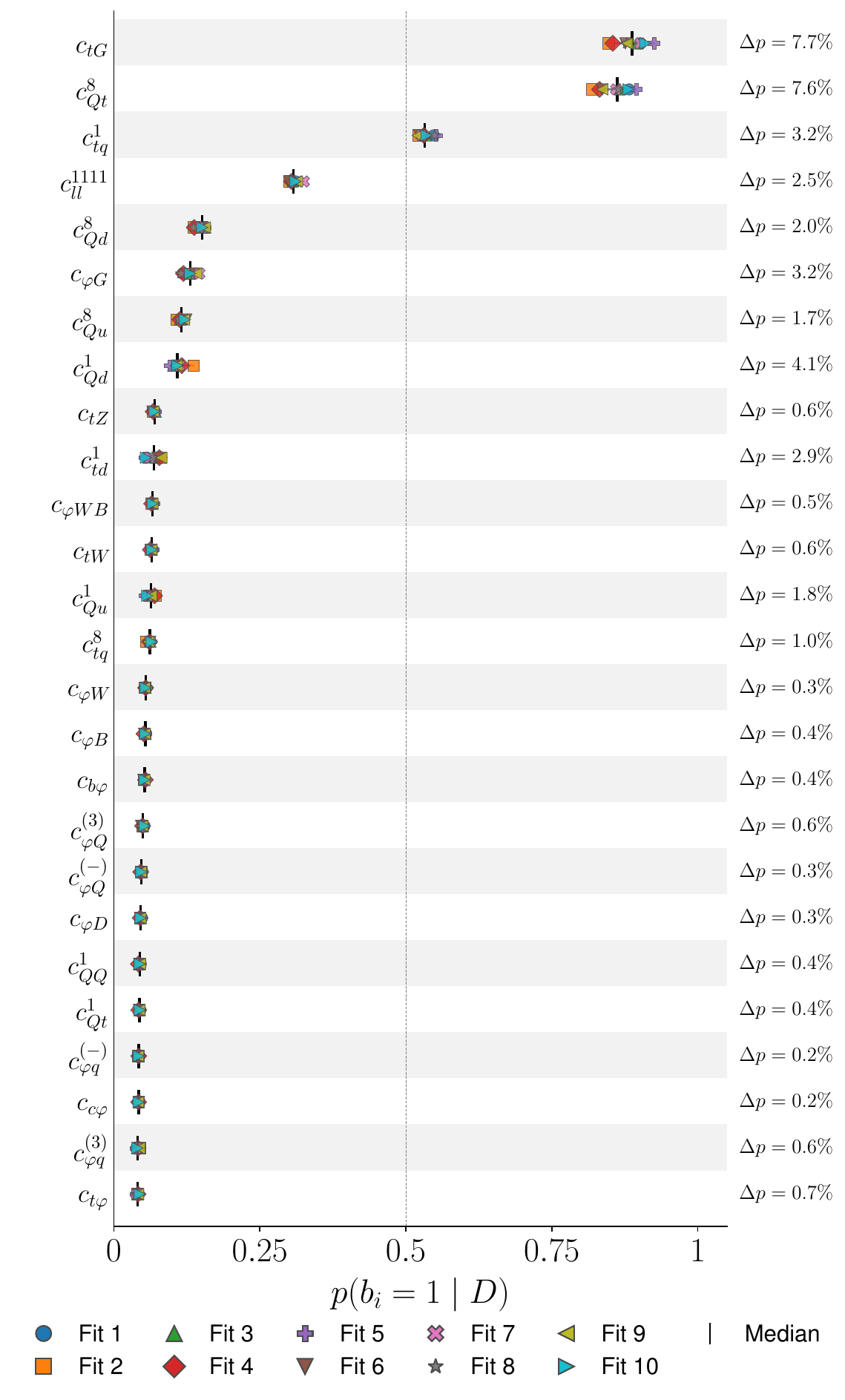}
    \caption{Marginal inclusion probability $p(b_i = 1 \mid D)$ obtained
    from ten independent GA runs differing only in random seed, for the top 26 operators ranked by median
    inclusion probability, at $\mu_0 = 10\,\mathrm{TeV}$ and linear EFT
    order. The black tick marks the median across runs; the grey line
    spans the full range. $\Delta p_i$, the spread across runs, is
    reported alongside each operator.}
    \label{fig:ga_stability1}
\end{figure}

\begin{figure}
    \centering
    \includegraphics[width=0.85\linewidth, page=2]{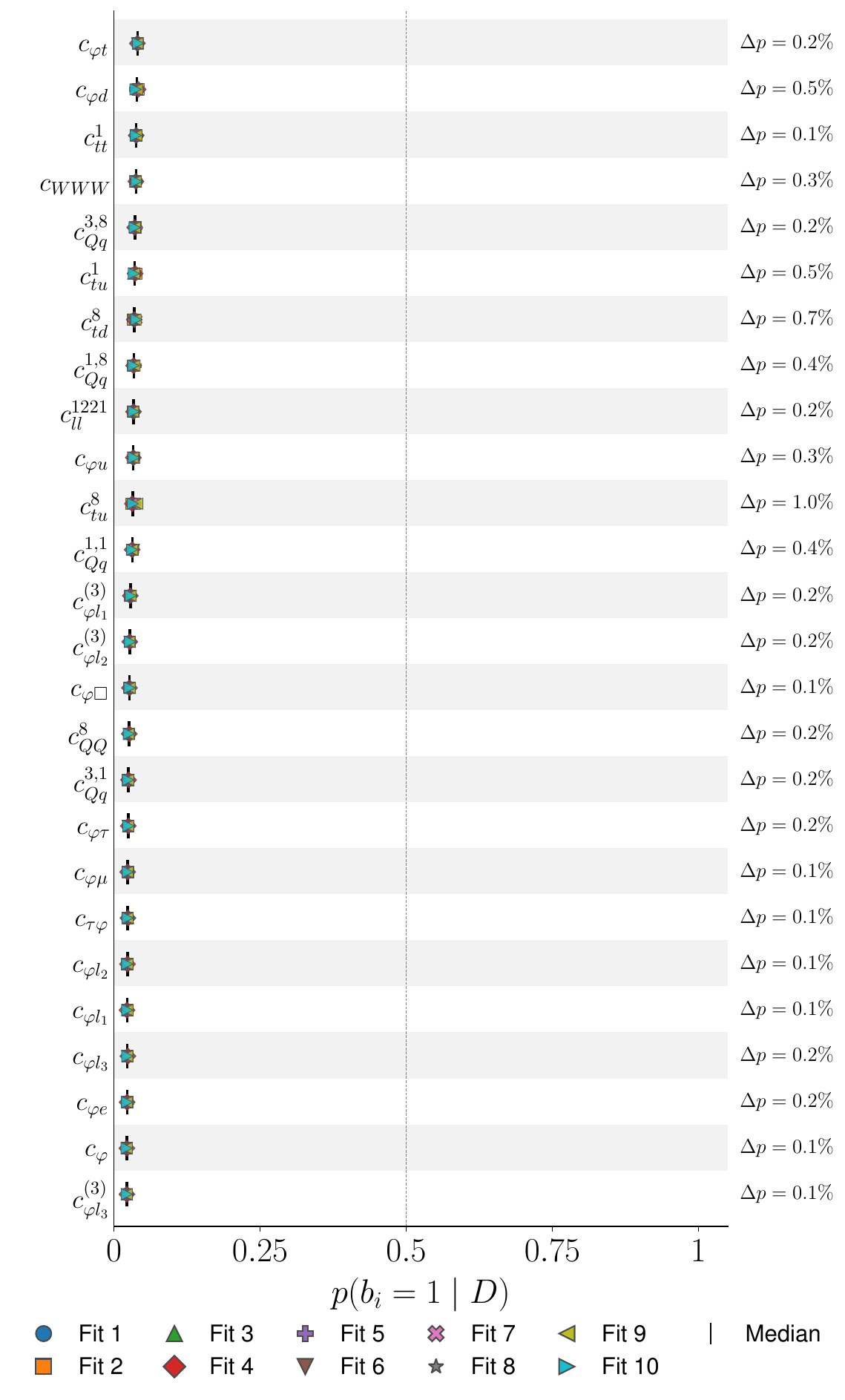}
    \caption{Same as Fig.~\ref{fig:ga_stability1}, for the remaining 26 operators.}
    \label{fig:ga_stability2}
\end{figure}

Figs.~\ref{fig:ga_stability1} and~\ref{fig:ga_stability2} show, for every
operator in the basis, the inclusion probability obtained in each of the
ten runs together with $\Delta p_i$. Across the full operator basis, we
find no operator
exceeding the $10\%$ threshold we adopt to flag instability. The two largest values are found for $c_{tG}$ ($\Delta p = 7.7\%$) and
$c^8_{Qt}$ ($\Delta p = 7.6\%$), the two operators identified in
Sect.~\ref{sec:inclusion} as reaching the highest marginal inclusion
probabilities ($\sim 90\%$) and, per Sect.~\ref{sec:correlations}, the
most strongly correlated pair in the posterior ($\phi_{ij} \simeq 0.6$).
Crucially, even for this pair the \emph{ranking} is unaffected by the
choice of seed: $c_{tG}$ is found to be more probable than $c^8_{Qt}$ in
every one of the ten runs, so the run-to-run variation reshuffles the
precise value of $p(b_i = 1 \mid D)$ without ever reshuffling the
conclusion drawn from it. A second tier of operators, including
$c^1_{Qd}$, $c^1_{tq}$, $c_{\varphi G}$, $c^1_{td}$, and $c^{1111}_{ll}$,
shows spreads of a few percent, while for the remaining operators spreads
are at or below the percent level.

The discovery statistic $p(\mathcal{H}_{\mathrm{SM}}|D)$ is equally 
reproducible across the same ten runs, with 
$p(\mathcal{H}_{\mathrm{SM}}|D) = (99.478 \pm 0.008)\%$ (mean and standard 
deviation).

As a further reproducibility check, we compare the ten best models returned 
by each of the ten independent GA runs on the real dataset. The three 
leading models discussed in Sect.~\ref{sec:global_evidence}, 
$\{c^1_{tq}, c^8_{Qt}, c_{tG}\}$, 
$\{c^1_{tq}, c^8_{Qt}, c^{1111}_{ll}, c_{tG}\}$, and 
$\{c^8_{Qt}, c_{tG}\}$, are recovered identically in all ten runs. Pooling 
across all ten runs, only twelve distinct operator subsets ever appear in a 
top-10 list, all lying within $\Delta\mathrm{BIC} \lesssim 2.5$ of the best 
model; each run's top-10 captures ten of these twelve, with the two left 
out varying from run to run. The set of statistically competitive models is 
thus stable across seeds. The ten models displayed in 
Fig.~\ref{fig:bayes_model} correspond to the ten lowest-BIC models 
identified by pooling the ten independent runs discussed above. We stress 
that this check establishes that the GA consistently samples from the same 
pool of competitive models across independent seeds; it does not by itself 
establish that this pool is complete.

Completeness is addressed by a complementary test, probing a distinct 
failure mode: whether the GA can miss an isolated low-dimensional model 
whose constituent operators are not individually competitive enough to 
survive selection long enough to be combined. We exhaustively evaluate 
every one-, two-, three-, and four-operator subset of the full 52-operator 
basis, together with the SM ($294{,}204$ models in total), at linear order 
and $\mu_0 = 10\,\mathrm{TeV}$. The twelve lowest-BIC models coincide 
exactly, in the same order and with matching BIC values, with the pool 
identified above: the GA misses none of the leading restricted models. 
Checking against the full archive of a single representative GA run, of 
the kind used to generate the results of Sect.~\ref{sec:eft_results}, the 
best-scoring model not visited is found at $\Delta\mathrm{BIC} = 3.17$; 
pooling the archives of all ten independent runs used in the 
reproducibility check above, this rises to $\Delta\mathrm{BIC} = 6.19$. We repeat 
this exhaustive scan at quadratic order and confirm that the ten models 
shown in the corresponding panel of Fig.~\ref{fig:bayes_model} coincide with the GA's own 
top-10; the best-scoring model not visited by the GA is found only at 
$\Delta\mathrm{BIC} = 7.33$. No isolated low-dimensional model anywhere in 
the full basis, at either EFT order, therefore outperforms or competes 
with the set already recovered by the GA.

Taken together, these results confirm that the population size and
number of generations adopted throughout this work are sufficient for
the GA-visited ensemble to give a stable approximation to the model
posterior: individual inclusion probabilities can shift by a few percent
between runs, but the resulting hierarchy of operator support, the
global discovery statistic, and the identity of the leading models are
all unchanged, and every physical conclusion drawn from them in this
work is unaffected. The complementary exhaustive check further confirms that this stable,
reproducible pool of models is not undercut by any missed low-dimensional
alternative: up to $k \leq 4$ active operators, every model overlooked by
the GA is uncompetitive with the pool it recovers, at either EFT order. Reducing the residual
run-to-run variation further, to the sub-percent level, is
straightforwardly achievable by increasing the population size or number
of generations at the cost of additional computing time, and represents
a simple avenue for tightening precision should a future analysis
require it.

\subsection{Comparison against exact enumeration at reduced dimensionality}
\label{sec:ga_exhaustive}

The previous reproducibility check establishes that
the GA-visited ensemble does not depend on the particular random seed, but
it cannot by itself distinguish a genuinely accurate ensemble from one that
consistently converges to the same biased region of model space. To close
this gap, we compare the GA against the exact posterior obtained by
exhaustive enumeration of the full model space at reduced operator
dimensionality, $d = 20$, for which all $2^{20} \approx 1.0 \times 10^6$
candidate models can be evaluated directly. The operator subset is chosen
to include heavy four-quark, dipoles, and electroweak
operators identified as physically relevant in Sect.~\ref{sec:global_evidence}. 
The GA is run with the
same population size and number of generations used throughout the paper,
so that it visits a comparably small fraction of the full model space ($\mathcal{O}(10k)$ hypotheses) as in
the $d = 52$ analysis, rather than approaching exhaustive coverage simply
because the space itself is smaller.
This reduction in dimensionality is deliberate. The
property under test is whether the GA's stochastic search is systematically
biased when sampling only a small fraction of model space, a property of the
search mechanism itself, and exhaustive enumeration is the only setting where
this can be checked against ground truth. Agreement with the exact posterior
constitutes evidence for the reliability of the search mechanism, complementing
the seed-reproducibility check.

\begin{figure}
    \centering
    \includegraphics[width=0.85\linewidth]{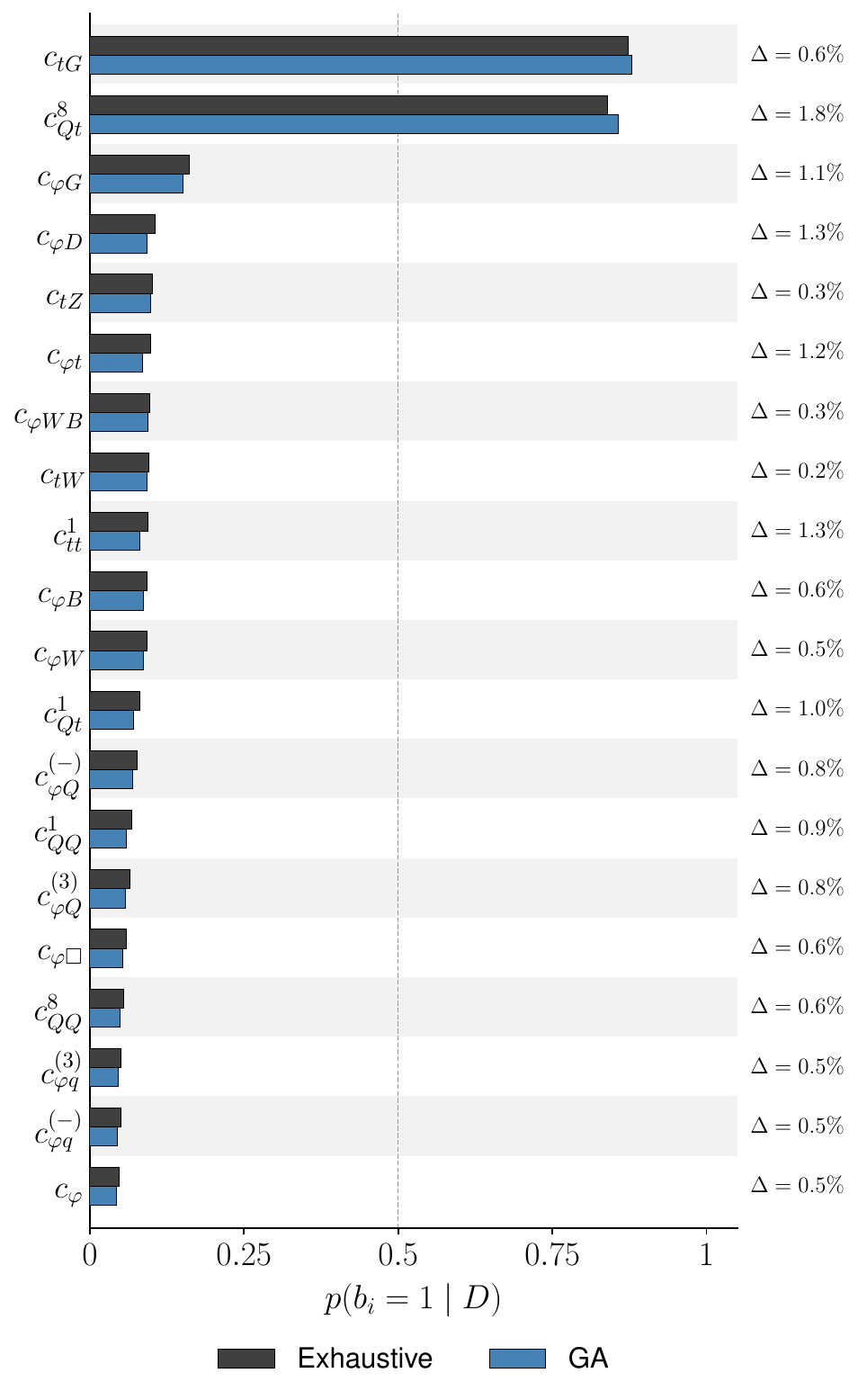}
    \caption{Marginal inclusion probability $p(b_i = 1 \mid D)$ obtained
    from exhaustive enumeration of the full $d = 20$ model space (dark
    grey) compared against the GA-visited ensemble using the same search
    hyperparameters adopted throughout this work (blue). Operators are
    ordered by decreasing exact inclusion probability; $\Delta$ denotes the
    absolute difference between the two.}
    \label{fig:ga_exhaustive}
\end{figure}

Fig.~\ref{fig:ga_exhaustive} compares the marginal inclusion probabilities
$p(b_i = 1 \mid D)$ obtained from the exact enumeration and from the GA for
each of the twenty operators. The agreement is close across the full basis:
the largest discrepancy is found for $c^8_{Qt}$, $\Delta = 1.8\%$, followed by
$c_{\varphi D}$ and $c^1_{tt}$, each at $\Delta = 1.3\%$;
the mean discrepancy across all twenty operators is $0.8\%$, and the
smallest, $\Delta = 0.2\%$, is found for $c_{tW}$, with $c_{tZ}$ and $c_{\varphi WB}$ at $0.3\%$.

We note that the GA-estimated probability is smaller than the exact value
for all operators except $c_{tG}$ and $c^8_{Qt}$, for which it is marginally
larger. The general trend is plausibly a consequence of the way
$p(b_i = 1 \mid D)$ is constructed as a ratio of sums over models: the
denominator is dominated by the handful of models near the global BIC
optimum, a region the GA is designed to locate reliably, while the numerator,
restricted to models containing a specific operator, draws more of its weight
from the broader set of sub-optimal variants that a finite search is less
likely to enumerate exhaustively. These differences are in any case small
compared to the separation in inclusion probability that distinguishes
operators favoured by the data from the bulk of the basis
(Sect.~\ref{sec:inclusion}), and are comparable to, or smaller than, the
run-to-run spread already documented in Sect.~\ref{app:ga_convergence}.

\clearpage

\bibliographystyle{JHEP}
\bibliography{references.bib}
\end{document}